\documentclass[11pt,preprint]{aastex61}

\newcommand       \AV        {A_{\rm V}}
\newcommand       \AB        {A_{\rm B}}

\newcommand       \Teff      {T_{\rm {eff}}}
\newcommand       \logg      {{\rm {log}}\,g}
\newcommand       \feh       {\rm{[Fe/H]}}

\newcommand       \Msun      {M_{\sun}}
\newcommand       \Ks        {K_{\rm S}}

\newcommand       \gps       {g_{\rm P1}}
\newcommand       \rps       {r_{\rm P1}}
\newcommand       \ips       {i_{\rm P1}}
\newcommand       \zps       {z_{\rm P1}}
\newcommand       \yps       {y_{\rm P1}}
\newcommand       \gs        {g^{\prime}}
\newcommand       \rs        {r^{\prime}}
\newcommand       \is        {i^{\prime}}
\newcommand       \Gbp       {G_{\rm BP}}
\newcommand       \Grp       {G_{\rm RP}}
\newcommand       \Rv        {R_{\rm V}}
\newcommand       \HI        {\rm H\,\uppercase\expandafter{\romannumeral1}}
\newcommand       \HII       {\rm H\,\uppercase\expandafter{\romannumeral2}}
\newcommand       \asil      {\rm \langle a \rangle_{sil}}
\newcommand       \agra      {\rm \langle a \rangle_{gra}}
\defcitealias{hz18}{Paper\,I}

\accepted{by ApJ February 12, 2020}

\begin{document}

\title{A Systematic Study of the dust of Galactic Supernova Remnants I. The Distance and
the Extinction}

\correspondingauthor{Biwei Jiang}
\email{bjiang@bnu.edu.cn}

\author{He Zhao}
\affiliation{Department of Astronomy, Beijing Normal University, Beijing 100875, P. R. China}
\affiliation{University C\^ote d'Azur, Observatory of the C\^ote d'Azur, CNRS, Lagrange
Laboratory, Observatory Bd, CS 34229, \\
06304 Nice cedex 4, France}
\email{he.zhao@oca.eu}

\author{Biwei Jiang}
\affiliation{Department of Astronomy, Beijing Normal University, Beijing 100875, P. R. China}

\author{Jun Li}
\affiliation{Department of Astronomy, Beijing Normal University, Beijing 100875, P. R. China}

\author{Bingqiu Chen}
\affiliation{South-Western Institute for Astronomy Research, Yunnan University, Kunming,
Yunnan 650091, P. R. China}

\author{Bin Yu}
\affiliation{Department of Astronomy, Beijing Normal University, Beijing 100875, P. R. China}

\author{Ye Wang}
\affiliation{Department of Astronomy, Beijing Normal University, Beijing 100875, P. R. China}

\begin{abstract}

By combining the photometric, spectroscopic, and astrometric information
of the stars in the sightline of SNRs, the distances to and
the extinctions of 32 Galactic supernova remnants (SNRs) are investigated. The stellar
atmospheric parameters are from the SDSS$-$DR14$/$APOGEE and LAMOST$-$DR5$/$LEGUE
spectroscopic surveys. The multi-band photometry, from optical to infrared,
are collected from the {\it Gaia}, APASS, Pan--STARRS1, 2MASS, and {\it WISE} surveys.
With the calibrated {\it Gaia} distances of individual stars, the distances to
15 of 32 SNRs are well determined from their produced extinction and association with molecular clouds.
The upper limits of distance are derived for 3 SNRs. The color
excess ratios $E(\gps-\lambda) / E(\gps-\rps)$ of 32 SNRs are calculated, and their
variation with wavebands is fitted by a simple dust model. The inferred
dust grain size distribution bifurcates: while the graphite grains have comparable
size to the average ISM dust, the silicate grains are generally larger. Along the
way, the average extinction law from optical to near-infrared of the Milky Way
is derived from the 1.3 million star sample and found to agree with the CCM89
law with $\Rv=3.15$.

\end{abstract}

\keywords{ISM: supernova remnants -- dust, extinction}

\section{Introduction} \label{sec:intro}	

Supernovae (SNe) have profound effects on their environments as a violent process,
by injecting kinetic energy on the order of $10^{51}$\,erg into the interstellar
medium (ISM). The high-speed expansion of the ejecta produces strong shocks
that interact with the surrounding circumstellar and interstellar dust. This
interaction destroys the ambient interstellar dust and alters the size of the dust 
grains \citep{Laki2015}. Whether the grains
become smaller or bigger is not clear. On one hand, the energetic particles
($v>150\,{\rm km\,s^{-1}}$) more often in fast shocks knock atoms off the grain
surface \citep[the so-called sputtering process,][]{Dwek96} which results in a
deficit of small grains in supernova remnants (SNRs). On the other hand, the
grain--grain collision dominant in slow shocks ($v \le 50-80\,{\rm km\,s^{-1}}$)
mainly destroys big grains which results in a deficit of big grains \citep{Jones94}.
\citet{noz07} modelled the process of dust evolution in SNRs and found that the
survival of dust in SNRs depends on the density of the ambient medium, e.g. in
the case of $\rm 1\,cm^{-3}$ typical of our ISM, small ($< 0.05\,\micron$) grains
are completely destroyed, intermediate (between $0.05-0.2\,\micron$) grains are
trapped into the dense shell, and big grains ($>0.2\,\micron$) are not changed
significantly. The increase of the ambient density would shift the border size
towards smaller. In addition to the initial size, the fate also depends on the
composition of the dust. \citet{noz07} found that silicate dust may be more easily
influenced than carbonaceous dust. Detection of SiO line emission in shocks is
evidence of destruction of silicate dust \citep{Guillet09}. Thus, the shocks can
modify the dust composition abundance in addition to the dust size distribution.

The changes in the dust size and component leave footprints in the dust emission
from the SNRs. \citet{Andersen11} analyzed the infrared (IR) spectral energy distribution (SED)
and spectra of 14 Galactic SNRs and derived the ratio of very small grains to big
grains. They
found this ratio to be higher for most of their SNRs than that in the plane of
the Milky Way, which may be explained by dust shattering that destroy big grains.
Meanwhile, this ratio is lower than in the Galactic plane for two SNRs, indicating
that sputtering is responsible for the dust destruction. \citet{TD13} modelled
the SED with the \emph{Herschel} far-infrared observation of the Crab Nebula and found
a fairly large maximum grain size up to $0.1\,\micron$ or bigger, although
\citet{Temin12} argued for a small grain size of $<0.05\,\micron$ from the dust
emission measured by \emph{Spitzer} at relatively short wavelengths. On the dust composition,
\citet{Andersen11} found all of their 14 SNRs show evidence of PAH emission, and
silicate emission at $20\,\micron$ in two SNRs. \citet{Arendt99} analyzed the
ISO/SWS spectrum of Cas\,A and identified an emission feature at $\sim$22\,$\micron$
which could not be fitted by typical astronomical silicate. This is confirmed
by the \emph{Spitzer}/IRS observation in which \citet{rho08} detected a feature around
$21\,\micron$ and which they explained by combining a few types of oxygen-bearing
dust species.

Similar to the dust emission, the extinction by dust reveals the changes
in the dust size distribution and composition of SNRs as well because the extinction
law depends sensitively on the size and composition of the dust. As an
experiment, \citet{hz18} (hereafter \citetalias{hz18}) studied the extinction
and dust of the Monoceros SNR (G205.5+0.5). By taking the stars along the
sightline of the SNR as tracers of the SNR extinction, \citetalias{hz18}
derived the distance to the Mon SNR and the extinction produced by the SNR
dust in the near-infrared (NIR) bands (namely the 2MASS $JH\Ks$). The present work
is an extension of \citetalias{hz18} whose principles and methods are followed.
Our goals include estimation of distances to the SNRs and study
of their extinction law to explore the effect of the SN explosion on both
the SN ejecta and mainly the interstellar dust grains. Firstly, the stars in the sightline
towards the SNR are selected, and their interstellar extinctions are calculated.
Secondly, the distance to the SNR is estimated from the position where the extinction
increases sharply due to the higher dust density of the SNR than its foreground ISM.
Last, the extinction law and the dust properties of the SNR are derived.
It should be clarified that the dust of SNR is merely produced by the SN ejecta.
The dust produced by the SN ejecta is hardly more than
a solar mass, which is shown by the case of young SNRs. For example,
\citet{Matsuura2015} derived $0.8\,\Msun$ for SN\,1987A,  \citet{bev17} derived
$\sim$1.1\,$\Msun$ for Cas\,A, and \citet{OB15} $0.6\,\Msun$ for the Crab Nebula.
However, the expansion of the SN shock waves can sweep up massive enough ambient
interstellar dust to produce
relative large extinction jumps, in particular so considering that many SN explosions occur in
large molecular clouds (MCs). \citet{Draine2009} estimated the ISM mass to be swept by a SN
explosion of about 1000\,$\Msun$. Compared to \citetalias{hz18}, we make a few
important
improvements: (1) the stellar distance is measured by {\it Gaia} and extracted
from the newly released {\it Gaia} DR2 catalog. The distance from the {\it Gaia}
parallax is a geometrical measurement without dependence on stellar model and
interstellar extinction; (2) the extinction curve is derived from optical to
IR, which reveals more accurately the properties of the SNR dust,
whereas \citetalias{hz18} only studied the NIR extinction law that might be
universal \citep{WJ14}; (3) the study is systematic by including
many more SNRs and the conclusion will be drawn from a wider range of SNRs.

The paper is organized as follows. In Section \ref{sec:data}, we describe
the data sets and objects used in this work and the quality control to build
the star sample. The methods underpinning the determination of intrinsic color indices
and extinctions are introduced in Section \ref{sec:method} in which an extinction--distance
model to fit the reddening profile at a given line of sight is also
described. The main results and discussions are presented in Section \ref{sec:result},
and summarized in Section \ref{sec:conclusion}.

\section{Data} \label{sec:data}

We collect stellar parameters (effective temperature $\Teff$, surface
gravity $\logg$, and metallicity $\feh$) from the LAMOST and APOGEE surveys,
photometric data from the {\it Gaia}, Pan--STARRS1, APASS, 2MASS, and {\it WISE}
surveys. The distance information is extracted from the {\it Gaia} parallax.
After the quality control, each star in our sample has full information of
stellar parameters, 11-band photometry from optical to IR, and distance.

\subsection{LAMOST} \label{subsec:lamost}

LAMOST (the Large sky Area Multi-Object fiber Spectroscopic Telescope) is a
reflecting Schmidt telescope operated by the National Astronomical
Observatories, Chinese Academy of Science (NAOC). It can obtain 4000 spectra
in a single exposure at the resolution $R=$\,1800 with the wavelength coverage
of $\rm 3700\,\AA < \lambda < 9100\,\AA$ \citep{zhao12,deng14}. LAMOST completed
its fifth-year survey in July 2017, and the LAMOST--DR5 catalog contains over
9 million spectra. In this work, we take use of its catalog of over 5 million
A, F, G and K-type stars for which the stellar parameters are highly
reliable.\footnote{See \url{http://dr5.lamost.org/}.}

\subsection{APOGEE} \label{subsec:apogee}

APOGEE (Apache Point Observatory Galaxy Evolution Experiment) is a
high-resolution ($R \approx$\,22,500), $H$-band (1.51--1.70\,$\micron$)
spectroscopic sky survey, providing accurate stellar parameters ($\Teff$,
$\logg$, and $\rm [M/H]$) for giant stars \citep{eis11}. The newly released
APOGEE data in SDSS--DR14\footnote{See \url{http://www.sdss.org/dr14/}}
\citep{Abolfathi18} includes both APOGEE--1 (SDSS--\uppercase\expandafter{\romannumeral3})
data and the first two years data from APOGEE--2
(SDSS--\uppercase\expandafter{\romannumeral4}).
The new data expand the previous APOGEE sample in both size (from $\sim$163,000 to
over 270,000 stars) and spatial coverage \citep{maj16}.

\subsection{Pan--STARRS1} \label{subsec:ps1}

The Pan--STARRS1 3$\pi$ survey was made with a 1.8\,m telescope on
Haleakala \citep{hod04}. The observations were conducted by using the 1.4
billion pixel GPC1 camera and performed in five broad filters: $\gps$,
$\rps$, $\ips$, $\zps$, and $\yps$, with effective wavelengths of $\rm
4800\,\AA$, $\rm 6200\,\AA$, $\rm 7500\,\AA$, $\rm 8700\,\AA$, and $\rm 9600\,\AA$,
respectively \citep{ona08}. The accuracy of relative and absolute photometry
is better than $1\%$ \citep{ton12,sch12}. The $\zps$, and $\yps$ data are
not used in our research because the APASS survey makes no observation in
these bands for relatively bright stars.

\subsection{APASS} \label{subsec:apass}

The AAVSO Photometric All-Sky Surey (APASS) is conducted in five filters:
Johnson $B$ and $V$, and Sloan $\gs$, $\rs$, $\is$, and reliable photometry
ranges from about 10.0\,mag to 17.0\,mag in $V$-band \citep{hm14}. APASS is
carried out to bridge the gap between the surveys, like Tycho--2 $B_T$ $V_T$
survey (accurate at $< 10$\,mag) and SDSS (saturating at $<$
14\,mag) \citep{mun14,hm14}. The APASS$-$DR9 covered about 99\% of the sky and
contains over 60 million objects\footnote{See \url{https://www.aavso.org/apass}}
\citep{hen16}. As the Pan--STARRS1 survey has a saturation problem with relatively
bright stars, we use APASS survey, instead of Pan--STARRS1, to provide the photometries
in $gri$-bands for stars brighter than 14.0, 14.4, and 14.4 mag
in the $\gps$, $\rps$, and $\ips$ bands respectively, following
the criteria of \citet{sch16}. To complete the transformation of photometric
system from Pan--STARRS1 to SDSS, the quadratic relations by \citet{ton12} are used.

\subsection{2MASS} \label{subsec:2mass}

The Two Micron All Sky Survey (2MASS) provides the most widely used photometric
data in the NIR bands: $J$ ($1.24\,\micron$), $H$ ($1.66\,\micron$),
and $\Ks$ ($2.16\,\micron$) \citep{coh03}. The completeness of 2MASS is over 99\%
for $J<15.8$, $H<15.1$, and $\Ks<14.3$\,mag, and the 2MASS point source catalog we
used in this work contains over 470 million objects \citep{cut03}.

\subsection{WISE} \label{subsec:wise}

{\it WISE} (Wide-field Infrared Survey Explorer) is an IR space telescope
with a diameter of 40\,cm. It was launched in 2009, and performed a mid-IR
full-sky survey in four bands: $W_1$ ($3.35\,\micron$), $W_2$ ($4.60\,\micron$),
$W_3$ ($11.56\,\micron$), and $W_4$ ($22.09\,\micron$) \citep{Wright10}.
The AllWISE source catalog provides the photometric data. We only take the
$W_1$ and $W_2$ bands into use because the sensitivities of the long-wavelength
bands $W_3$ and $W_4$ are significantly lower and unable to match the sensitivities of
the other bands.

\subsection{Gaia} \label{subsec:gaia}

The newly released {\it Gaia} DR2 contains 1.3 billion sources with
trigonometirc parallaxes, three bands photometry ($\Gbp$, $G$, $\Grp$), and
proper motions \citep{gaia2018a}. The central wavelengths of $\Gbp$, $G$,
and $\Grp$ are 532, 673, and 797\,nm, respectively \citep{Jordi2010}. The
distances used in our work are derived from the {\it Gaia} parallaxes and
calibrated by \citet{Bailer-Jones18}.

\subsection{Data Combination and Quality Control} \label{subsec:data}

With these many data sets into use, we first combine the LAMOST and APOGEE
catalog, where the stellar parameters from APOGEE are kept for their higher
precision for the overlapping sources although these two sets of stellar
parameters are generally consistent within the error range \citep{Anguiano18}.
Then the stars with stellar parameters from the spectroscopic surveys are
cross-matched with the multi-band photometric data from {\it Gaia}, Pan--STARRS1,
APASS, 2MASS, and {\it WISE}, as well as {\it Gaia} distances. All the catalogs are
cross-matched within 1$\arcsec$.

The measurements of stellar parameter, photometric magnitude and parallax
are required to fulfill the following requirements in order to obtain
reliable results:

\begin{enumerate}
  \item The photometric error in all bands is smaller than 0.1\,mag.
  \item The errors of stellar effective temperature and surface gravity are
        $\sigma_{\rm \Teff}<150$\,K, $\sigma_{\rm \logg}<0.2$\,dex respectively.
  \item Sources from LAMOST should have snr\,$g>30$ (signal-to-noise ratio in the
        $g$-band), and sources from APOGEE should have ${\rm S/N}>100$ and $\rm
        VSCATTER
        <0.3$\,km/s (the velocity scattering of multi-epoch measurements) to exclude
        binary stars.
  \item The fractional error of stellar distance from {\it Gaia} DR2 is smaller than
      30\%,
        and the sources with failed calibration are excluded.
\end{enumerate}

Furthermore, the dwarfs and giants are judged according to the following criteria:

\begin{enumerate}
  \item Dwarfs:  $\logg>4$ for 4000\,K $<\Teff<$ 6500\,K; $\logg>3.5$ for 6500\,K
      $<\Teff<$ 8500\,K.
  \item Giants: 3900\,K $<\Teff<$ 5400\,K, $1<\logg<3$.
  \item For both dwarfs and giants, $-1.0<\feh<0.5$ is required because the measured
  metallicity has large uncertainty outside this range.
\end{enumerate}

The latest data sets extend significantly the ranges of stellar parameters for both
dwarfs
and giants in comparison with the star sample used in \citetalias{hz18}.
Finally, 1,115,536 dwarfs and 221,820 giants are selected to constitute our
star sample, all with stellar parameters, {\it Gaia} distance, and 11-band
photometries: $\gps$, $\Gbp$, $\rps$, $G$, $\ips$, $\Grp$, $J$, $H$,
$\Ks$, $W_1$, $W_2$. Stars can be as faint as $\gps = 20$\,mag and trace the
dust with $E(\gps-\Ks) \approx 14.0$\,mag. Meanwhile, their distances can
reach the zone farther than 15\,kpc, though mostly within 6\,kpc. These
characteristics make it possible to detect Galactic SNRs in an extensive
range, and perform a systematic study of their distances and extinction laws.

\section{Method} \label{sec:method}

\subsection{Intrinsic Color Indices: from Optical to IR}
\label{subsec:ici}

The `blue edge' method is widely used to derive the stellar intrinsic color
index from the effective temperature $\Teff$ for a given luminosity class
\citep[see][]{duc01,WJ14,xue16,jian17}. The premise is that the
extinction-free stars constitute the blue edge in the
$\Teff - C_{\lambda_1 \lambda_2}$ (observed color)
diagram for large stellar surveys, because they have the smallest observed
colors which are indeed their intrinsic colors at the given $\Teff$. By deriving
the analytic function of this blue edge, the stellar intrinsic color can
then be calculated from its effective temperature. In practice, the median
color of some fraction of the bluest stars is taken as the intrinsic color
instead of choosing the exact bluest one, because the photometric and
parameter's uncertainties would shift the colors in some range.
\citet{jian17} found that 10\% is an appropriate fraction for the sources
with photometric errors around 0.05\,mag, i.e. the median color of the
bluest 10\% stars can represent the intrinsic color for each assigned
temperature interval. In comparison with the NIR bands,
metallicity has heavier influence on intrinsic colors in optical bands
involved in the present work. So the star sample is further divided into six
groups with a step of 0.25\,dex in the measured $\feh$ from $-1$ to $0.5$. With
this subdivision, the sample stars for $\gps-\Gbp$
and $\gps-G$ with $-1<\feh<-0.75$ are not adequate in number, and the bluest
20\% rather than 10\% stars are chosen. The
comparison between different $\feh$ groups finds that the intrinsic colors of dwarfs
are hardly affected by metallicity. Meanwhile giant stars show a systematic
change with $\feh$, i.e. their intrinsic colors increase with $\feh$.

An exponential function is fitted to the relations between the intrinsic colors
and $\Teff$ in each group:

\begin{equation}\label{eq:IciFit}
C^0_{\lambda_1\lambda_2} = A_0~{\rm exp}\left(\frac{\Teff}{A_1}\right) + A_2.
\end{equation}

\noindent For the 11 photometric bands, 10 intrinsic color indices are
derived with $\gps$ as the reference band, namely $(\gps-\Gbp)_0$,
$(\gps-\rps)_0$, $(\gps-G)_0$, $(\gps-\ips)_0$, $(\gps-\Grp)_0$,
$(\gps-J)_0$, $(\gps-H)_0$, and $(\gps-\Ks)_0$, $(\gps-W_1)_0$,
$(\gps-W_2)_0$ for each metallicity group and for dwarfs and giants
respectively. The coefficients are presented in Table \ref{tab:coeICI}.
The case of color index $(\gps-\Ks)_0$ is shown as an example in Figure
\ref{fig:ici}, as well as the comparison of different $\feh$ groups.

As discussed in \citetalias{hz18}, the uncertainty brought by the fitting
technique is on the order of several thousandths of a magnitude. The error induced by
the
adopted bluest fraction is 0.03\,mag for dwarfs and 0.06\,mag for giants \citep{jian17}.
Since we further divided the sample according to $\feh$, the influence of $\feh$
is reduced in comparison with \citetalias{hz18}. In conclusion, the
total uncertainty of the calculated intrinsic color is comparable to the photometric
error ($\sim$0.05\,mag) for most of the sample stars.

\subsection{The extinction--distance model} \label{subsec:adMod}

With the intrinsic color indices of individual stars derived from the
blue-edge method, their color excesses are calculated straightforwardly by
subtracting the intrinsic colors from the observed. Then the variation
of reddening along distance towards a given line of sight can be obtained
with the help of {\it Gaia} distance of individual stars. In order to determine
the distance to a SNR, the extinction--distance model used in \citet{chen17}
is adopted because this model is insensitive to the outliers:

\begin{equation}\label{eq:total-ext}
A(d) = A^0(d) + A^1(d),
\end{equation}

\noindent where $A(d)$ is the total extinction measured along the sightline.
$A^1(d)$ is the contribution from the dust in the SNR. Assuming that the
integrated extinction is dominated by one SNR cloud, $A^1(d)$ can be
described by a function,

\begin{equation}\label{eq:ext-dist}
A^1(d) = \frac{\delta A}{2} \times \left[1+{\rm erf}\left(\frac{x-d_0}{\sqrt{2}\,\delta
d}\right)\right]
\end{equation}

\noindent where $\delta A$ represents the amplitude of the extinction jump,
i.e. the extinction of the SNR, $d_0$ is the distance to the center of the
SNR and $\delta d$ is the width of the SNR calculated from the angular
diameter and $d_0$ of the SNR.

$A^0(d)$ represents the extinction provided by the diffuse interstellar dust.
\citet{chen17} suggested a two-order polynomial to describe $A^0(d)$, which is
reasonable for the nearby zone, but the integrated reddening will drop quickly with distance at large. 
To avoid this unreasonable tendency, we replace the quadratic term with
a square root,

\begin{equation}\label{eq:ism-profile}
A^0(d) = a \times d + b \times \sqrt{d},
\end{equation}

\noindent where $a$ and $b$ are the fitting coefficients. It should be kept
in mind that how the interstellar reddening changes with distance
is still not clear and needs further investigation. Here we only present a
mathematical fitting to the foreground and background extinction.

A Markov Chain Monte Carlo (MCMC) procedure \citep{emcee13} is performed to
optimize the free parameters in the model. We apply a scheme similar to
\citet{Kos17} to make the procedure more stable. A first Markov chain with
100 walkers and 200 steps is run to estimate the initial parameters for the
final chain. Then 1000 steps with 128 walkers are run, and the last 750
steps from each walker are used to sample the final posterior. The median
values (50th percentile) of the posterior distribution are taken as the
best estimates, with uncertainties derived from the 16th and 84th percentile
values. If the parameters tend to gather into two distinct groups during the
MCMC sampling, a new distance component is added to the model, which means
there are two distances for one sightline. 6 SNRs are assigned with a second
distance component.

The distance to the Rosette Nebula (a well studied $\HII$ region) is fitted
as a test of the model. \citetalias{hz18} determined its distance to be
1.55\,kpc, which coincides with literature results. 298 dwarfs and 43 giants
are selected from our star sample towards the region of the Rosette Nebula
defined by \citetalias{hz18}. The fitting results are shown in Figure
\ref{fig:Rosette} for all the 10 color excesses. The derived distances from
different bands are highly consistent with each other with a
standard deviation of 0.06\,kpc. Their mean value, 1.58\,kpc, agrees very well with
previous works. Figure \ref{fig:corner} is the corner plot of the
parameters at $E(\gps-\Ks)$, which shows all the one and two dimensional
projections of the posterior distributions of the parameters. The blue
squares and lines demonstrate the best-fit values of the parameters, which
also exhibits the order of the uncertainties of the fitted parameters.

\subsection{Foreground extinction} \label{subsec:foreground}

To measure the extinction produced by the SNR alone, foreground extinction must
be extracted. In Eq.\ref{eq:total-ext}, the foreground extinction is described
by $A^0(d)$ (Eq.\ref{eq:ism-profile}). In the case that the observed data can be
well fitted by Eq.\ref{eq:total-ext},
the foreground extinction can be determined simultaneously with the extinction
of the SNR ($\delta A$). Unfortunately, some cases cannot be well fitted by
Eq.\ref{eq:total-ext} mostly because the sample stars are inadequate in number
either due to the small region of the SNR or the incomplete observation. In
such case, we have to select a model for foreground interstellar extinction.
We choose a typical diffuse sightline to represent the common extinction--distance
model of ISM to simplify the calculation, although the extinction--distance
model should change with sightlines. This may introduce additional errors in
estimating the extinction of these SNRs, but it has little influence on the
distance determination.

\citet{wang17} reported a very diffuse region centered at ($l=165\fdg0,\,b=0\fdg0$,
hereafter named ``$l165$'') based on the Milky Way's CO emission map \citep{Dame01}.
With a square region of $2\arcdeg \times 2\arcdeg$ towards ``$l165$'', 320 stars
are picked from our star sample. Although this region was proven to be diffuse,
the increase of extinction with distance still becomes steeper beyond 3\,kpc
as seen in Figure \ref{fig:CI}, which is caused by the Perseus arm. We have to
apply the extinction--distance model (Eq.\ref{eq:total-ext}) to ``$l165$'', which
resulted in both the ISM extinction and the extinction produced by the
Perseus arm around 3.5\,kpc. The derived ISM component is then used to represent
the foreground extinction for the SNRs whose reddening profiles are not well 
fitted by the model Eq.\ref{eq:total-ext}.

\subsection{The SNR region} \label{subsec:region}

The projected region for every SNR needs to be defined carefully to choose
appropriate sample stars because the SNR morphology is usually irregular.
The centers and angular sizes reported in the revised catalog of 294 Galactic
SNRs by \citet{Green19} provide a very good initial estimation of the regions
(the referred regions).
We take a simple circle with the major radius from this catalog as the referred
regions. The regions are further refined according to the radio observations
of the SNRs.

The radio data are mostly taken from the Effelsberg 100\,m telescope
observation \citep{reich90,reich97}. The images of G74.0--8.5, G82.2+5.3,
and G89.0+4.7 come from the Effelsberg Medium Latitude Survey (EMLS) at 1.4\,GHz
\citep{uyaniker99}. Some objects lack the Effelsberg observation and alternative
data are taken. For G156.2+5.7, G178.2--4.2 and G182.4+4.3, the observations
are made by the Urumqi 25\,m telescope at 6\,cm \citep{gao10}. The data of
G65.3+5.7 and G70.0--21.5 are from the 4850\,MHz GB6S survey \citep{condon91,condon94}.
For G159.6+7.3, the 325\,MHz observation from the Westerbork
Northern Sky Survey \citep[WENSS][]{rengelink97} is used.

For G65.1+0.6 and G108.2--0.6, only giant stars in the selected regions are
available. We doubled the size of the regions to enclose some foreground
dwarf stars to determine the foreground extinction. The neighbouring intensive
radio sources blur the borders of G67.6+0.9, G108.2--0.6, G152.4--2.1, and
G213.0--0.6. G65.3+5.7, G70.0--21.5, and G159.6+7.3 are very faint and large
SNRs. The radiation field of some SNRs, e.g.\ G109.1--1.0, G119.5+10.2, and G189.1+3.0,
extend beyond the referred regions, so larger polygonal fields are applied. If a
SNR cannot be distinguished from its radio observation, the referred region is
applied. The projected region of G205.5+00.5 (Monoceros) is from \citetalias{hz18}.

With the defined regions, we complete the source selection towards the sightlines
of SNRs from our star sample. For 45 SNRs, there are dwarf and/or giant stars from our
selected
sample. But some SNRs in the Galactic center direction have too few stars in the
sightline,
they are excluded since no reliable results can be derived. Finally, 32 SNRs are
selected to study their distances and extinctions. The results are
presented in Figure \ref{fig:snrReg}, along with the SNR regions and radio observations.
To make the figures concise, we omit their contour lines and colorbars. Different figures
are in different intensity scales. It should be noted that the applied irregular
regions prefer to follow some level of the contour lines enclosing the SNRs.

\section{Results and Discussions} \label{sec:result}

\subsection{The SNR distance} \label{subsec:snrDist}

There are 32 Galactic SNRs in total covered by our star sample. We obtain reddening 
profiles for 21 of them by our extinction--distance model. Their distances are 
calculated by taking the average of the fitted distances from the reddenings of 
5 IR bands relative to $\gps$, i.e. $\gps-J$, $\gps-H$, $\gps-\Ks$, $\gps-W_1$, 
and $\gps-W_2$. The visual bands are dropped because their much smaller color 
excess relative to $\gps$, $E(\gps-\lambda)$, will introduce significant uncertainties. 
Meanwhile, the dispersion of distance from different bands is taken into account 
for estimating the final uncertainty together with the error provided by the MCMC 
procedure. In addition to these 21 SNRs, the distance of G78.2+2.1 can be reliably 
recognized by the conspicuous extinction jump although the extinction--distance 
model cannot fit its reddening profile because of the absence of distant tracers. 
For other SNRs, the analysis yields upper limits or null results of distance.
7 of them have only giant stars as the distance and extinction tracers so that the
background extinction cannot be determined. Consequently no accurate distance is
determined, instead the upper limit is constrained by comparing the extinction
of tracers with the reddening profile of ``$l165$''. The run of reddenings traced 
by our star sample towards every SNR plus the ``$l165$'' are shown in Figure 
\ref{fig:snrDist}, and their distances are presented in Table \ref{tab:snrDist1}--\ref{tab:snrDist3}, 
as well as the results from the literatures. It should be noted that the distances
indicated in the figures may be slightly different from those reported in the tables
because we only present the fitting results at $E(\gps-\Ks)$ in Figure \ref{fig:snrDist}
while Table \ref{tab:snrDist1}--\ref{tab:snrDist3} results from the average distance
from five bands.

The 32 SNRs are classified into three levels according to the credibilities of 
the results. Level A includes 15 SNRs with well determined distance. Because 
the SNRs are all located in the Galactic plane full of MCs, the detected extinction 
jump can plausibly be caused by the MCs in the sightline. However, since the 
core-collapse SN originates from massive stars, these SNRs are expected to associate
with MCs. Then the distance of such SNR should agree with that of the associated 
molecular cloud. This argument is supported by the work of \citet{YuB2019} which 
takes the distance to the MC at the sightline as that of the SNR. Therefore, we 
searched for the SNRs in our sample which were previously suggested to be associated 
with MCs. It is found that 9 of them have been verified to be associated with 
MCs by \citet{JiangB2010} and the references therein, namely G78.2+2.1, G89.0+4.7, 
G94.0+1.0, G109.1--1.0, G166.0+4.3, G189.1+3.0, G190.9--2.2, G205.5+0.5, and 
G213.0--0.6. In addition, G152.4--2.1, G160.9+2.6 and G182.4+4.3 are suggested 
to have associated MCs by \citet{YuB2019}. Thus these SNRs are more likely to 
sweep up massive amounts of dust of the ambient MC that can cause apparent 
extinction jump along the sightline. One more support comes from the reddening 
profiles derived from \citet{Green2019} which show no other apparent extinction 
jumps within 7\,kpc towards these SNRs (see Figure \ref{fig:comGreen}), i.e. no 
confusing jump exists. Therefore, the distances to these 12 SNRs are regarded to 
be well determined. G93.7--0.2, G156.2+5.7 and G206.9+2.3 are also classified as
Level A because of their apparent extinction hikes and their dust properties 
consistent with the expectation from SNRs (see Section \ref{subsubsec:mrn}).

The measurements of 7 SNRs in Level B have lower credibilities. Their distances 
are either upper limits or less restricted. Both of G82.2+5.3 and G108.2--0.6 
exhibit apparent extinction jump, but they are still classified as Level B because 
no associated MC has been reported up to now and the detected extinction jump may 
be caused by the foreground or background MC, though there is still high possibility 
that the distance is correct with further investigation of the association of 
the SNR and MC. For these sources, reliable upper limits of distance are determined
for G74.0--8.5, G113.0+0.2 and G127.1+0.5, while not enough tracers are available 
towards G65.3+5.7 and G116.9+0.2 to well constrain the fittings. Level C contains 
5 estimates with great uncertainties and 5 failed cases. Among them, the explorations 
to G179.0+2.6 and G180.0--1.7 are suggested to only detect some foreground clouds. 
Detailed discussions for individual cases are presented in Appendix \ref{appsec:snrDist}.
The following discussions concentrate mainly on 22 SNRs in Level A (15) and B (7).

\subsubsection{Comparison with the distances from two dust maps}
\label{subsubsec:distCompare}

The reddenings of SNRs can also be inferred from three-dimensional (3D) dust maps.
Based on a recently released 3D dust map of \citet{ChenBQ2019}, \citet{YuB2019}
analyzed the dust mappings of 12 SNRs towards the Galactic anti-center. 
Although they claimed reliable determination for only four sources, 
the distance is successfully determined for seven SNRs, namely G152.4--2.1,
G160.9+2.6, G182.4+4.3, G189.1+3.0, G190.9--2.2, G205.5+0.5 and G213.0--0.6. All
of their results (presented in bold face in Table \ref{tab:snrDist1}) are highly
consistent with ours, which enhances the reliability of both results. This
consistency is remarkable. Basically, \citet{YuB2019} determined the distance of
the MC overlapping with the SNR in the radio image, which can be the distance of
the SNR only if the SNR and the MC are interacting. Meanwhile, we determined the
distance of the SNR according to the stars in the sightline of the SNR other than
the stars to the neighbouring MC. The agreement between the two
distances implies that the SNR and the MC are interacting for the 7
cases. On the other hand, \citet{YuB2019} did not find any associated MCs
for the 5 other SNRs and thus no distance was determined. Although massive stars
are expected to be associated with MCs, many OB stars are field stars
and some SNRs may be unassociated with any MC. In addition, the SNR from type 
Ia SN usually have no association with MC.

\citet{Green2019} constructed a 3D map of the Milky Way dust reddening with 800
million stars observed by Pan--STARRS1, 2MASS, and {\it Gaia}. The typical angular
resolution of the map is from $3\farcm4$ to $13\farcm7$, and the distance
ranges from 63\,pc to 63\,kpc. To retrieve a representative reddening profile
from the \citet{Green2019} map for each of the 32 SNRs as well as the Rosette
Nebula, we select the sightline of the measured high extinction region within the
SNR, and sample the dust map from 100\,pc to 7\,kpc with a step of 100\,pc. It is
noted that for Monoceros SNR, the sightline to a MC centered at $(l=204\fdg107,\,b=0\fdg
471)$ reported by \citet{SuY2017} is applied. Our results are compared with this 
reddening map in Figure \ref{fig:comGreen}, where the fluctuation of the reddening 
profile is caused by the uncertainty of the dust map. High consistency is achieved 
for most of the SNRs in Level A, see e.g. G78.2+2.1 and G190.9--2.2. In a few cases, 
like G89.0+4.7, no apparent extinction jump is present in the \citet{Green2019} map, 
which may be due to the relatively low spatial resolution of the \citet{Green2019} map 
that smoothed out the feature. For SNRs G82.2+5.3 and G108.2--0.6 there are additional
jumps from the dust map besides the one revealed by our analysis with one distance
component. The sharp increase at $\sim$0.3\,kpc and the platform afterwards towards
G178.2--4.2 (Level C) supports our estimate of distance $< 0.3$\,kpc. Moreover, our
upper limits of distances for SNRs in Level B are consistent with this dust map as well.
For G116.5+1.1, the position ($\sim$0.7\,kpc) of the extinction jump is highly
consistent with our nearest tracer (0.68\,kpc), so we also suggest 0.68\,kpc
as a possible distance of this SNR, noted in Table \ref{tab:snrDist3}.

Because the LAMOST and APOGEE spectroscopic surveys are limited by stellar
brightness, our star sample cannot provide the reddening profile with high
spatial resolution. Therefore for most of the cases, we can only derive
the extinction jump for the whole SNR, rather than inferring a two-dimensional
extinction map. If the dust in the SNR leads to a significant
extinction jump along the sightline, accurate measurements can be made. While
for the SNRs with low extinction and within
a complex interstellar environment, like G180.0--1.7, we cannot select specific
fields like \citet{chen17} to reveal the SNR behind a huge foreground MC.
3D extinction maps help us revise our results. On the other hand, the
stellar parameters from spectroscopy determine more accurately stellar reddening
than the photometric colors. In the future, an expansion of the sample and an
improvement of the method can be expected in combination with 3D dust maps.

\subsubsection{Comparison with distances measured by other methods}
\label{subsubsec:dist-comp}

Besides the 3D extinction analysis, some other methods are applied to
estimate the distances to SNRs. First, the kinematic method based on the $\HI$ and/or
CO absorption lines and the rotation curve of the Milky Way is widely used. 
This method suffers the ambiguity problem in the inner disk and the
large uncertainty in the outer disk.
Second, the distance to SNR can be inferred from the associated objects with
known distances, such as OB star \citep{Humphreys78}, MC \citep{GF07}, or pulsars
\citep{Kramer03}. The uncertainty of this method comes from both the identification 
of the association and the distance of the associated
objects. Third, extinction measurements based on red clump (RC) stars
can  trace the distances \citep[e.g.][]{dur06,shan18}. Fourth, the empirical
relation between surface brightness ($\Sigma$) and physical diameter ($D$) of
SNRs ($\Sigma-D$ relation) is often adopted to obtain SNR distance as well. 
The dispersion of the relation brings about uncertainty and the
faintness of the object would bear more uncertainty.
Fifth, the Sedov estimate can be made for some shell-type SNRs with 
X-ray observation for which the inhomogeneity of the ISM complicates the result.
Others include the calculation of distance using proper motion \citep{boumis04}
or shock velocity \citep{Blair05}. The error analysis will be addressed in more 
details for individual object in Appendix \ref{appsec:snrDist}.

We found about 53 measurements in the literature for 22 SNRs in Level A (15) and B
(7) with 7 methods (6 methods mentioned above plus \citet{YuB2019}).
The comparison between our estimates and the distances in the literature is shown
in Table \ref{tab:snrDist1} and \ref{tab:snrDist2}, and Figure \ref{fig:dist-comp} where
different methods are in different colors and
illustrated in the legends (``Others'' indicates 3 works based on proper motion
or shock velocity). The comparison with \citet{YuB2019} has been discussed in
Section \ref{subsubsec:distCompare}. Figure \ref{fig:dist-comp} visually exhibits the
large dispersion between literature measurements as well as the large uncertainty in
many cases. So
there is no possibility for our measurements to coincide with all previous results. On
the contrary, discrepancy is expected.

Kinematic method is most widely used which accounts for over
a third of the gross (18/53). Good agreements can be found for some cases like
G109.1--1.0, G189.1+3.0, and G190.9--2.2, while large discrepancies exist as well.
In the direction of Galactic anti-centre, kinematic method tends to gain larger
distances which may be caused by the deviation due to non-circular motion \citep{TL2012}
or the larger uncertainty of the rotation curve in the outer disk in comparison with
inner disk due to invalidity of the tangent velocity method. The kinematic method
suffers the well-known ambiguity problem as well.
10 estimates derived from the associated objects show the difference with our results
on a similar level to the kinematic method.  Distance measurements based on RC stars are mainly
applicable
to the SNRs in the first and fourth quadrants of the Milky Way because RC stars are
much less abundant towards Galactic anti-centre. So only 3 of 22 SNRs are
studied with the RC stars. Two of them (G82.2+5.3 and G109.1--1.0) yielded much
larger distances than ours as well as other results, and the estimate to G89.0+4.7 is
comparable to ours. Distances derived from the $\Sigma-D$ relation (7 cases) are
much closer to our estimates compared to the results with methods of kinematics
and associated objects.  The 5 Sedov estimates all yield distances larger than
ours.  The 3 studies based on proper motion or shock velocity
provide some informative restrictions on the SNRs in Level B whose distances are not
well determined in this study. The detailed analysis of the differences can be found in Appendix \ref{appsec:snrDist}.

\subsubsection{Distribution in the Galactic plane} \label{subsubsec:snr-distribution}

With the measured distances, Figure \ref{fig:snr-distribution} shows the spatial
distribution of 22 SNRs in Level A and B projected on the Galactic plane. Due
to the limits of
the APOGEE and LAMOST spectroscopic surveys, most of the SNRs under investigation
are in the second and third quadrants of the Milky Way. It can be seen that the SNRs
are mainly located in the Local arm. 6 SNRs are suggested to reside in the Perseus arm.
Three cases, i.e. G89.0+4.7, G93.7--0.2, and G94.0+1.0, are found between the two arms.

\subsection{The extinction curve} \label{subsec:snrExtcurve}

The extinction curves vary from one sightline to another, revealing the
wavelength-dependent extinction law in different environments. As the absolute
extinction is usually difficult to measure and uncertain, we use the color
excess ratios (CERs), i.e. $E(\gps-\lambda) / E(\gps-\rps)$, to represent
the extinction curves for the sightlines to the 32 SNRs, as well as the
``$l165$'' and the Rosette Nebula. For SNRs whose reddening profiles were
well model-fitted, the color excess is then $\delta A$ in Eq.\ref{eq:ext-dist}
-- the amplitude of the extinction jump at a given color. For the other SNRs,
we firstly calculate the color excesses for individual sample stars in and
behind the SNRs by subtracting the foreground ISM contribution according to the
reddening profiles of ``$l165$'', and then derive the CERs by proportional
fitting to the color excesses $E(\gps-\lambda)$ and $E(\gps-\rps)$. Because the
stars towards G70.0--21.5 and G159.6+7.3 show lower extinctions than the
sightline of ``$l165$'', no subtraction is performed for these two sightlines, which
can bring about uncertainties in the subsequent results. 

To characterize the properties of SNR dust, the CCM89 formula \citep{CCM89} and 
a simple dust model are applied
to fit the extinction curves. In total, from $\Gbp$ to {\it WISE}$/W_2$, we
have 10 CERs for the fittings. But \citet{Wang2019} reported significant amount
of curvature of CERs (the deviation from the linear relationship between
two color excesses for highly reddened stars) for the {\it Gaia} bands for heavily reddened stars because of their
broad bandwidths. Since the correction for the curvature needs stellar
intrinsic flux distribution that is unavailable, $G$ and $\Gbp$ are removed in further
fitting, while $\Grp$ is usable due to its longer wavelength and much less affected.
Because the APASS and Pan--STARRS1 filters cover the wavelength range of {\it Gaia},
the removal of the two {\it Gaia} bands should not change the result. Therefore,
the fitting is performed to the other 8 CERs.

\subsubsection{CCM89} \label{subsubsec:CCM89}

\citet{CCM89} presented a one-parameter model (CCM89) to approximate the various
extinction curves, which received wide application. CCM89 is characterized by the ratio
of the total extinction to the selective extinction $\Rv = \AV/E(B-V) = \AV/(\AB-\AV)$.
With the derived CERs, $\Rv$ is sampled from 1 to 10 with a step of 0.01 to find
the best-fit for each SNR. A Monte Carlo simulation based on the values and errors
of the CERs determines the uncertainty of $\Rv$. The best-fit $\Rv$ values are shown
in Table \ref{tab:ext-fitting} together with the error.

For 18 cases, $\Rv$ value ranges between 3.1 and 5.5, while for 10 other sightlines $\Rv
< 3.1$.
There are 4 SNRs with $\Rv > 5.5$: $\Rv = 5.98$ for G213.6--0.6, $\Rv=6.44$ for
G108.2--0.6,
$\Rv=6.61$ for G166.0+4.3, and the largest $\Rv$ of 7.71 for G179.0+2.6. But for
G179.0+2.6, its CERs show a dramatic rise in the $\ips$ and $\Grp$ bands, following a
platform in IR bands, indeed the CCM89 formula cannot reasonably characterize this
queer curve, and the resultant $\Rv$ of 7.71 is unreliable.

\subsubsection{The simple dust model} \label{subsubsec:mrn}

CCM89 is a mathematical formula, which cannot tell the dust properties, and as shown
above, fails in some cases. Besides, CCM89 is derived mainly from the extinction
curves in the UV and optical bands, significant deviations appear in the IR bands,
which occurs in our fitting. Such deviation can also be seen from comparing the CCM89
law and the WD01 \citep{WD01} model in Figure \ref{fig:wd01} at $\Rv=3.1$, 4.0 and 5.5
respectively. The comparison shows that CCM89 yields a smaller $\Rv$ value for a given
extinction curve, and the differences increase with $\Rv$. In order to better fit the
derived extinction as a function of waveband, we build a simple dust model.

Because the derived extinction curves do not have enough observational points or
wavelength
coverage to constrain the relatively numerous parameters of WD01, a simpler dust model
is applied to the extinction curves. This model considers two main
species of the dust in ISM: graphite and silicate. To simplify the model, their mass
ratio is fixed as $\rm M_{sil}/M_{gra} = 2:1$. The size distributions of both graphite
and silicate grains follow the MRN model \citep{MRN77}, i.e. $\rm n(a) \propto
a^{-\alpha}$.
The model is fixed by the power law index of the grain size distribution
$\alpha_{\rm gra}$ and $\alpha_{\rm sil}$ for graphite and silicate grains respectively.
For convenience of comparison, we defined
the average size of dust grains by

\begin{equation}\label{eq:ave-size}
\rm \langle a \rangle = \frac{\int_{a_{min}}^{a_{max}}{a \cdot
n(a)\,da}}{\int_{a_{min}}^{a_{max}}{n(a)\,da}},
\end{equation}

\noindent where $\rm a_{min}$ and $\rm a_{max}$ are $0.005\,\micron$ and $0.25\,\micron$
respectively. The fitting results are shown in Table \ref{tab:ext-fitting} and
Figure \ref{fig:mrn-fit}. As only the indexes for the dust size distribution power law 
are adjustable, the fitting fails for 7 SNRs. All of the other
25 SNRs yield larger $\Rv$ values than that by CCM89, and generally,
their differences increase with $\Rv$ (see Figure \ref{fig:mrn-ccm}), which
resembles the situation of the difference between the CCM89 formula and the WD01
model.

The following discussions are based only on the 22 SNRs in Level A/B with
reliable distance measurements while those in Level C are not included. The fitted $\Rv$ value
is $> 3.23$ (the value for
the diffuse ``$l165$'' region) for 13 SNRs, and $> 4.50$ for 6 SNRs.
Generally, the value of $\Rv$ increases with the grain size because the larger
grains is more efficient in extinction at longer wavelength. Figure
\ref{fig:mrn-com2} displays the $\Rv$ values from our dust model as a function of the average radii
of the graphite ($\agra$) and silicate ($\asil$) grains respectively.
All the SNRs except G108.2--0.6 have the average radius of graphite grains
$\agra$\,$\sim$\,0.008\,$\micron$.
In contrast, $\asil$
varies from 0.007 to 0.16\,$\micron$, mostly much larger than $\agra$.
In comparison, the Rosette Nebula and the ``$l165$'' region have $\asil$ very
close to their $\agra$, both are either $0.008\,\micron$ or $0.009\,\micron$.
If the dust towards the ``$l165$'' region with $\Rv=3.23$ is typical for diffuse
ISM, it may be concluded that the average size of the silicate grains in
SNRs is bigger than that in diffuse ISM, although the graphite grains
are comparable to those in diffuse ISM. This large size of silicate grains
leads to the larger $\Rv$ values since silicate grains constitute two-thirds of
the dust in the model and thus are dominant contributor to the extinction. The model
of \citet{noz07} suggested that silicate dust is more easily destroyed by SN
explosion than carbonaceous dust. The apparent increase of $\asil$ in some SNRs
may be explained by the destruction of small silicate grains by the sputtering
process. Together with the detection of SiO line emission in shocks \citep{Guillet09},
the large size of silicate grains supports the idea that the very small silicate
grains may be destroyed during the SN explosion. It is true that
much larger $\asil$ than $\agra$ is obtained for the Level A SNRs.
On the other hand, the ratio of the graphite to silicate grain size could diagnose 
whether the detected extinction jump is caused by the SNR such as in the cases of  
G93.7--0.2, G156.2+5.7 and
G206.9+2.3 (Section \ref{subsec:snrDist}), although the distance is still uncertain
because of the limitation of our simple dust model and the influence of the ambient
ISM in particular for the large SNRs. Meanwhile, some SNRs in Level B with apparent
extinction jumps have similar radius of
$\agra$ and $\asil$ that is comparable to the diffuse ISM, e.g.
G82.2+5.3, which may question the credibility that the extinction is caused by the SNR.
G108.2--0.6 is a special case with
the largest $\agra$ (0.025\,$\micron$) comparable to its $\asil$ (0.022\,$\micron$).
It may be argued that what the $\Rv$ value reflects is the property of the dust
in the MCs where the SNRs reside. This possibility is difficult to rule
out. But since we are confined to the SNR region other than the whole MC,
the dust, even from the MCs, should have been processed by the SN explosion.
They may be called the SN processed dust in MCs. Moreover, the normal extinction curve
of the MCs can usually be represented well by the CCM89 curve, while the extinction curves derived here
usually deviates apparently from the CCM89 curve. These dust should have been influenced
by the SN explosion to some extent.

If we assume a similar ambient ISM environment for all the SNRs, which is certainly
very rough, the diameter is then positively correlated with the age of the SNR.
In order to know if $\Rv$ depends on evolution of the SNR, the distribution of
$\Rv$ with the SNR diameter ($D$) is shown in Figure \ref{fig:mrn-com3}, where $D$
is calculated from the distance ($d$) and the major angular diameter ($\theta$), $D=d
\times {\rm sin}(\theta)$. It can be seen that the largest values of $\Rv$ occur
only for small SNRs that have a diameter less than or around 20\,pc. According
to \citet{dra11}, the radius of a SNR under typical interstellar conditions ranges
from $\sim$5\,pc to 24\,pc during the Sedov--Taylor phase, thus these relatively
smaller SNRs are in the  early Sedov--Taylor phase. In principle, large $\Rv$
implies large grains, which may be explained by destruction of small particles by
the shock wave. On the other hand, there are quite a few SNRs with small $\Rv$,
which cannot be explained by this mechanism. The larger SNRs with $D > 40$\,pc
lack very large $\Rv$ values but mostly around 4.5, systematically higher than
the mean value for diffuse ISM. These large SNRs should be relatively old and thin
thus the results may suffer more uncertainty than the small SNRs. With age, the
SNR expands and sweeps up more ambient interstellar dust which would erase away
the features of the SN ejecta. In addition, the shock wave slows down with age
and its influence on the ambient dust weakens. Consequently, the dust in the SNRs
would look more like the ambient interstellar dust. However, as we discussed
in Section \ref{sec:intro}, the effect of SN explosion on the grain size
is complex, depending on the type of the SN explosion, the environment, the
dust species and grain size etc. We will study further in detail the
grain size distributions and the causes in a forthcoming paper.

\subsection{The average extinction law of the Milky Way} \label{subsec:ave-curve}

The large star sample used in this work contains over 1.3 million sources.
They have a wide spatial distribution, with declination from --10$\arcdeg$ to
60$\arcdeg$ and distance beyond 15\,kpc (mainly within 6\,kpc). These stars
cover various environments, including both very diffuse ISM and dense dust
clouds with extinction as high as $E(\gps-\Ks) \approx 14.0$\,mag. Therefore, the
extinction law derived from this sample shall be very representative for
the average ISM environment.

Firstly, the color excesses for all the stars at all the 10 bands, i.e.
$E(\gps-\lambda)$, are computed. Considering the photometric errors
and the uncertainty of intrinsic colors, sources with color excesses
smaller than --0.1 are dropped, and then
a linear fitting is applied to derived CERs with $E(\gps-\rps)$ as the
reference color excess. Apparent curvatures appear in the color excess--color
excess diagrams (Figure \ref{fig:ave-cer}) for heavily reddened stars in the
{\it Gaia} bands which are discussed in detail by \citet{Wang2019}. Thus the 3
{\it Gaia} bands are not used in deriving the average extinction law. Since the
Pan--STARRS1 filters have covered the {\it Gaia} waveband range, the results should
not be influenced. The CCM89 law fitting to the CERs results in
$\Rv = 3.15$, very close to the average values suggested by other works
\citep{dra03,SF11}. While our simple dust model yields a larger $\Rv = 3.42$
as demonstrated earlier. In addition, the average radii of graphite and silicate
grains are $\agra = 0.0082\,\micron$, $\asil = 0.0087\,\micron$ respectively,
the same as the sizes towards the ``$l165$'' sightline (cf.\ Table
\ref{tab:ext-fitting}),
which confirms that the ``$l165$'' sightline is a typical diffuse ISM \citep{wang17}.

\citet{Wang2019} suggested a modified CCM89 formula to trace the average
extinction law. To fit the extinctions derived by using RC stars as standard
candle, they re-determined the coefficients of the CCM89 formula. In their
equations, the coefficients of high-order terms are neglected which yielded
a smoother curve than CCM89. In Figure \ref{fig:ext-curve}, the average CERs
(red squares) together with the dust model result (red line)
derived here are compared with the traditional CCM89 law (black line) and the
newly modified law by \citet{Wang2019} (blue line) with $\Rv=3.1$. In general,
our law coincides with CCM89 from optical to NIR though becomes slightly flatter
in the NIR. On the other hand, the IR extinction curve deviates from that of
\citet{Wang2019}. This deviation is difficult to understand since both methods
seem to calculate the color excesses accurately by subtracting the intrinsic
color derived from spectroscopy. One possible reason is that we use various
types of stars to trace the extinction while \citet{Wang2019} used only red
clump stars although this should make no difference in deriving the extinction
law.

\section{Summary} \label{sec:conclusion}

This work conducted a systematic study of the distances to and the extinctions of
Galactic SNRs from the change of extinction with distance of the
stars towards the sightline of the SNR. Compared to \citetalias{hz18}, the
methods are improved in two aspects. One is that the metallicity effect is taken
into account besides effective temperature and surface gravity when stellar intrinsic
colors are determined. This improvement leads to more accurate intrinsic colors
and correspondingly color excesses. The other is that stellar distance comes from
the {\it Gaia} DR2 that shuns the uncertainties from the stellar model.

Incorporating the available data from the APOGEE and LAMOST spectroscopic
surveys, the distance and the extinction from optical to NIR are studied for
32 Galactic SNRs. The distances to 14 SNRs in Level A are well determined by our
extinction--distance model. The measured distance to G78.2+2.1 is
reliable because of its conspicuous extinction jump and classified as Level A
object as well. In addition, the upper limits of distance to 3 SNRs are derived. In
total, the distance estimation is provided for 22 SNRs mostly in the second and third
quadrants.
Our estimated distances are in general
very consistent with the results derived from the 3D dust maps by \citet{YuB2019}
and \citet{Green2019}. Most of the investigated SNRs are located in the Local
arm, with distances $<2.0$\,kpc.

The extinctions of the SNRs expressed by the color excesses $E(\gps-\lambda)$
are either derived from the extinction--distance model or by taking the mean
values of the extinctions of tracer stars. The wavebands used include the
{\it Gaia}/$\Gbp$, $G$, and $\Grp$, APASS/$gri$ and Pan--STARRS1/$gri$, 2MASS/$JH\Ks$,
and {\it WISE}/$W_1$ and $W_2$ filters. The changes with waveband of the color excess
ratios $E(\gps-\lambda) / E(\gps-\rps)$ are fitted by the CCM89 law and the
deviation is apparent in some cases, in particular in the IR bands. A simple
dust model is built to fit the observed CERs. It is found that the
values of $\Rv$ range from about 2.4 to 6.7. The size of the SNR dust grains inferred
from the model
is bifurcated -- while the graphite grains have an average radius around
0.008\,$\micron$
comparable to the case in the diffuse ISM, the silicate grains are generally large
with an average radius up to about 0.02--0.03\,$\micron$ which is significantly bigger
than the grain size in the diffuse ISM. Although this phenomenon may be explained
by that the silicate grains are more easily destroyed by the SN explosion, a
careful study is needed to understand the reasons.

This work has two by-products. One is the relation of stellar intrinsic colors
with effective temperature for both dwarf and giant stars in the bands from optical
to NIR. The other is the average extinction law of our Galaxy from 1.3 million stars,
which agrees with the CCM89 with $\Rv=3.15$ though slightly deviates in the NIR bands.

In the following work, we will investigate the dust properties by a more refined
model, and the dust mass of the SNRs from the extinction and dust properties.
Further, the influence of SN explosion on the interstellar dust will be
discussed.

\begin{acknowledgements}
We would like to thank Dr. Jacco van Loon for his very careful
reading of the paper and his comments and suggestions that helped to improve the paper
effectively. We thank Drs Haibo Yuan and Mengfei Zhang for their useful
suggestions.
This work is supported by the National Natural Science Foundation of China through the
project NSFC 11533002,
and the China Scholarship Council (No.201806040200).
This work has made use of data from the surveys by SDSS, LAMOST, {\it Gaia}, APASS,
Pan-STARRS, 2MASS and {\it WISE}.
\end{acknowledgements}

\software{astropy \citep{Astropy13},
          emcee   \citep{emcee13}}

\appendix

\section{Discussion on the distances to the individual SNR} \label{appsec:snrDist}

\subsection{Level A} \label{appsubsec:levelA}

\begin{itemize}
  \item{G78.2+2.1} is a nearby shell-type SNR, with two bright unresolved shells
  in the north and south \citep{Gao11}. From the $\HI$ absorption spectra,
  \citet{Leahy2013} determined the association of G78.2+2.1 and the $\HII$ region
  $\gamma$ Cygni Nebula overlapping with the southern shell, and the distance to the
  SNR is then 1.7--2.6\,kpc. An earlier study of another associated $\HII$ region
  IC 1318b suggested a distance of $1.5 \pm 0.4$\,kpc \citep{Landecker1980}.
  There is no tracer further than 1.2\,kpc in our star sample so that the model
  fitting failed. Nevertheless, the tracers between 0.8 to 1.1\,kpc display a very
  conspicuous extinction jump, which puts strict constraint on its distance. We
  suggest the average distance of these tracers, 0.98\,kpc to be the distance to
  G78.2+2.1. Within the uncertainty range, this distance coincides with that
  of \citet{Landecker1980} which adopted the distance of the $\HII$ region IC 1318b.

  \item{G89.0+4.7} (HB21) was discovered by \citet{BH53} at 159\,MHz. HB21 has
  an irregular shell structure at radio wavelengths, with a well-defined outer
  boundary \citep{Tatematsu90,Uyaniker03,Kothes06,Gao11}. Based on the $\Sigma - D$
  relation, the distance of HB21 was estimated to be 1.0--1.6\,kpc
  \citep{Willis73,CC76,CL79,Milne79}. By observing the brightest stars in the
  Cyg OB complex with which HB21 is associated, and assuming a maximum visual
  luminosity for red supergiants, \citet{Humphreys78} estimated the distance
  to be $800\pm70$\,pc. \citet{shan18} got a distance of $1.9^{+0.3}_{-0.2}$\,kpc
  based on RCs. Our estimation yields a distance being the largest
  since a conspicuous extinction jump occurs at 2.6\,kpc. This is
  mainly because there are not so many tracers within 1.5--2.5\,kpc. Additionally, the
  reddening profile derived from the dust map of \citet{Green2019} presents an
  increase of extinction from $\sim$2.0\,kpc (see the corresponding subpanel
  in Figure \ref{fig:comGreen}). Thus the distance of HB21 is suggested to
  be 2.3\,kpc (mean value of 2.0--2.6\,kpc) with an error of 0.3\,kpc, in rough
  agreement with the \citet{shan18} result.

  \item{G93.7--0.2} is a shell-type SNR with some diffuse extensions \citep{Gao11}.
  \citet{Uyaniker02} reported the neutral material around G93.7--0.2, and the
  $\HII$ region S124 in the lower left corner, which is visible in the corresponding
  subpanel of Figure \ref{fig:snrReg}. With the measurement of surrounding $\HII$
  velocity and the standard Galactic rotation curve, \citet{Uyaniker02} placed
  the SNR at a distance of $1.5 \pm 0.2$\,kpc. Our fitting shows a further distance
  at $2.16 \pm 0.02$\,kpc. Foreground stars present a higher run of reddening than
  that of ``$l165$''. G93.7--0.2 is classified as Level A because of its conspicuous 
  extinction jump and the dust grain radius consistent with a SNR, i.e. $\asil > \agra$.

  \item{G94.0+1.0}. \citet{FR03} derived a distance--velocity relation with the $\HI$
  21\,cm spectral line, and used this relation to measure the distance to any
  object with a known systematic velocity. With this method, \citet{Foster05}
  suggested that G94.0+1.0 is expanding within a stellar wind bubble at a distance
  of 4.5\,kpc. Many giant stars in our sample have apparently high extinctions in
  comparison with the reddening profile of ``$l165$'', which implies a distance
  $< 3$\,kpc. We suggest a distance of 2.53\,kpc, with a relatively large uncertainty
  of 1.08\,kpc. G94.0+1.0 was proposed to be a Perseus arm object
  \citep{Foster05,Kothes05}, and then it should be closer than 3\,kpc.

  \item{G109.1--1.0} (CTB 109). CTB 109 hosts an anomalous
  X-ray pulsar 1E 2259+586 \citep{Koyama89}. With the $\HI$ observations, the
  distance of CTB 109 is determined as 3.0\,kpc \citep{Kothes02} and $4.0 \pm
  0.8$\,kpc \citep{Tian10}, respectively. Discrepantly, \citet{dur06} derived
  a much larger distance of $\sim$6\,kpc with RC stars. In this study, there are
  not many sample stars towards
  G109.1--1.0, but they still lead to a reliable distance determination at $2.79
  \pm 0.04$\,kpc which is consistent with the kinematic estimate by
  \citet{Kothes02}.

  \item{G152.4--2.1} was identified as a SNR by \citet{Foster13}. Based on the
  $\HI$ emission and the Galactic rotation curve \citep{Foster2010}, the SNR is
  suggested to be within the Local arm at a distance of $1.1 \pm 0.1$\,kpc.
  Our fitting results in a smaller distance at $0.59 \pm 0.09$\,kpc. The
  discrepancy between the kinematic estimate and ours may be due to the application
  of an undisturbed circular rotation curve \citep{Foster13} which would introduce
  considerable uncertainty for directions of Galactic anti-centre like G152.4-2.1.
  On the other hand, \citet{YuB2019} suggested that G152.4-2.1 possibly associates
  with a MC located within 1.0 kpc, supporting our result. Meanwhile, the
  extinction jump
  around 1.5\,kpc may refer to a background MC.

  \item{G156.2+5.7} is the first Galactic SNR discovered by strong
  X-ray emission \citep{Pfe91}. In the radio bands, the SNR displays a large shell
  structure with low surface brightness \citep{Xu07}. Based on the H$\alpha$ images
  and low resolution optical spectra, \citet{GF07} suggested there is physical
  interaction between G156.2+5.7 and a clump of interstellar clouds. According
  to the distances to these clouds, the SNR distance is suggested to be $\sim$0.3\,kpc. 
  Other estimates are 1--3\,kpc by the observation of associated
  $\HI$ emission \citep{Reich92}; 1.3\,kpc \citep{Yamauchi99} and 3\,kpc \citep{Pfe91}
  from the analysis of X-ray data. Both \citet{Pfe91} and \citet{Yamauchi99} applied
  Sedov estimation to G156.2+5.7. The discrepancy (1.3\,kpc and 3\,kpc) between their results is mainly
  caused by their different assumptions of the local environment based on the
  X-ray observations. Our sample stars bring about a reliable fitting towards
  G156.2+5.7, which indicates a distance of 0.68\,kpc. The discrepancies
  between our result and the distances in literature are mainly due to the complex
  interstellar environments along this sightline. Tracers with high extinctions
  can be found from $\sim$0.5 to $>1.0$\,kpc.

  \item{G160.9+2.6} (HB9) is a large nearby SNR that has been mapped in multiple
  radio bands \citep{Gao11}. The distance to HB9 is determined by a small extinction
  jump at 0.54\,kpc from our sample stars. It is consistent with 0.6\,kpc by
  \citet{YuB2019}
  that proves the association of G160.9+2.6 with a MC. Considering the
  uncertainties, the measured distance is also consistent with the result of $0.8
  \pm 0.4$\,kpc based on the $\HI$ observations \citep{LT07}. A series of previous
  studies estimate the distance to be around 1\,kpc, such as $\ge 1.1$\,kpc
  \citep{Loz81}, 1.1\,kpc \citep{Leahy87}, and 1.3--1.8\,kpc \citep{CL79}.
  They show differences not only with ours but with others and with each
  other. These relatively old observations may suffer lower sensitivity and resultant
  larger uncertainties.

  \item{G166.0+4.3}. The unusual shape of SNR G166.0+4.3, two shells with
      significantly
  different radii, caused a series of studies \citep[see][]{Kothes06,Gao11}, while
  the only credible estimate of its distance seems to be $4.5 \pm 1.5$\,kpc by
  \citet{Landecker89} based on the $\HI$ observation. We obtain two possible
  distances: $0.88 \pm 0.01$\,kpc and $3.24 \pm 0.03$\,kpc respectively.
  The larger distance is favored and the closer jump is caused by some nearby cloud or
  the incompleteness of the tracers. An apparent jump of $E(\gps-\Ks)$ can be seen for
  stars further than 3\,kpc, but there is no spatial separation for nearby and
  distant tracers. A distance of 3.24\,kpc will make the SNR 240\,pc above the
  Galactic plane, which is slightly larger than the typical scale-height of dust
  \citep{DS01,Misiriotis2006}.

  \item{G182.4+4.3}. \citet{Kothes98} identified G182.4+4.3, and reported a distance
  of $\ge 3$\,kpc by radio observations. But a tiny extinction jump indicates
  a distance of $1.05 \pm 0.24$\,kpc that is consistent with the estimate
  ($\sim$1.1kpc) based on the
  associated MC \citep{YuB2019}. It should be mentioned that the distance of
  \citet{Kothes98} is very much model dependent.  A possible cause is  their
  underestimation of the ambient number density, i.e. a very small
  density of ISM (0.013\,$\rm cm^{-3}$) which leads the SNR to be very big and very
  far. While \citet{YuB2019}
  suggested that the SNR is associated with a MC so that the SNR is located in a dense
  medium and can sweep enough medium at much closer distance.

  \item{G189.1+3.0} (IC 443) is a well known SNR located in a dense cloud \citep{Troja2008},
  with distinct shells of various radii \citep{Lee08}. A lot of studies are made on IC 443 
  through out the whole spectrum, from gamma ray to radio
  \citep[e.g.][]{Fesen1984,WS03,Acero2016,ZhangS2018,koo16,Gao11,Planck2016}.
  It interacts with the nearby MC, and the CO observation revealed
  a half molecular ring structure surrounding the SNR, from east to north
  \citep{SuY2014}. An IR bright shell is also found in north west \citep{koo16}. 1.5\,kpc
  is the most widely quoted distance to IC 443, assuming that the SNR interacts
  with the $\HII$ region S249 \citep{FK80,Fesen1984} whose distance is determined
  by the parallaxes of some stars from the Gem OB1 association \citep{Cornett77}.
  This is confirmed by the empirical $\Sigma - D$ relation \citep{Milne79,CL79}
  and the SNR model \citep{Chevalier99,WS03}. From the optical systemic velocities,
  \citet{Loz81} placed IC 443 between 0.7 and 1.5\,kpc. A recent study of the
  kinematics of the north-eastern region of the SNR suggested a distance of 1.9\,kpc
  \citep{Ambrocio-Cruz2017}.
  Our fitting yields a distance of $1.80 \pm 0.05$ with an apparent extinction jump,
  which is consistent with previous studies.

  \item{G190.9--2.2}. \citet{Foster13} first identified G190.9--2.2 as a SNR, with
  a barrel-shaped structure like G152.4--2.1. Based on the $\HI$ observation,
  \citet{Foster13} put G190.9--2.3 at 1.0\,kpc, with an uncertainty of 0.3\,kpc.
  This result is consistent with our measurement $1.03 \pm 0.01$\,kpc. The sharp
  increases of extinction at all the 10 bands make the estimation very accurate.
  The shape of the fitted reddening profile is similar to G189.1+3.0 in that
  intense increase occurs behind the SNR.

  \item{G205.5+0.5} (Monoceros Nebula) is an old \citep[$1.5 \times 10^5$\,yr][]{gra82}
  SNR with fine filamentary structure \citep{dav63}. The distance to Monoceros was
  estimated to be around 1.5\,kpc by previous works \citep{dav78,gra82,lea86}.
  In \citetalias{hz18}, we reported a distance of 1.98\,kpc to Monoceros, and
  1.55\,kpc to Rosette Nebula. As we discussed in Section \ref{subsec:adMod},
  the distance of the Rosette Nebula is 1.58\,kpc with the recently released
  data and the new method, which is consistent with \citetalias{hz18}. While
  for Monoceros, the present work yields two distances: $1.13 \pm 0.01$\,kpc
  and $2.57 \pm 0.02$\,kpc respectively. The larger distance matches the position
  of the Perseus arm, while the small distance is consistent with the recent
  determination by \citet{YuB2019} with the assumption that Monoceros interacts with
  neighbouring MC. The large difference between present work and \citetalias{hz18}
  is mainly due to the method. In \citetalias{hz18}, we selected a reference
  region besides Monoceros to subtract the reddening contributed by foreground
  dust. The large dispersion of extinction from the reference regions shadowed
  the nearby extinction jump. Then only the tracers around 1.98\,kpc are
  outstanding. High extinction tracers around 2.0\,kpc can also be found in the
  corresponding subpanel in Figure \ref{fig:snrDist}. But our model preferred
  a more stable estimate, 2.57\,kpc, for the second cloud rather than the
  distance of the closest tracers. Similar to the S147 SNR, the confusion of
  the Monoceros SNR case should at least partly be attributed to its low
  extinction due to its old age and thus thin dust shell.

  The relation between the Monoceros SNR and the Rosette Nebula is widely debated.
  \citet{dav78} suggested that there is interaction between Monoceros
  and the Rosette Nebula, which is supported by \citet{xiao12} with the $\HI$ channel
  maps. Based on the optical, CO, and radio observations, \citet{SuY2017} determined
  that the MCs at $V_{\rm LSR} \sim 5\,{\rm km\,s^{-1}}$ and $\sim 19\,{\rm
  km\,s^{-1}}$
  are both physically associated with Monoceros. As the Rosette Nebula is also
  surrounded by the 3--12\,${\rm km\,s^{-1}}$ and 18--23\,${\rm km\,s^{-1}}$ MCs,
  they suggested an association between Monoceros and the Rosette Nebula. Furthermore,
  based on the dust map of \citet{Green15}, a region centered at $(l=204\fdg
  107,\,b=0\fdg 471)$, the CO peak emission of the $5\,{\rm km\,s^{-1}}$ MCs, was
  selected to derive a distance of 1.6\,kpc to the SNR which is consistent with
  the position of the Rosette Nebula. But an updated dust map of \citet{Green2019}
  indicates an extinction jump $<1$\,kpc at $(l=204\fdg 107,\,b=0\fdg 471)$ (see
  the corresponding subpanel in Figure \ref{fig:comGreen}) consistent
  with our estimation. While the Rosette Nebula is still kept at $\sim$1.6\,kpc (the
  last subpanel in Figure \ref{fig:comGreen}) the Monoceros SNR
  is $\sim$450\,pc in front of the Rosette Nebula, they  can hardly  interact
  with each other. The associated MCs to Monoceros and Rosette Nebula may be located
  at
  different distances although they have similar velocities, for example,
  the MCs associated with NGC 2264 at about 900\,pc also have $V_{\rm LSR}
  \sim 5\,{\rm km\,s^{-1}}$ \citep{SuY2017}.

  \item{G206.9+2.3}. \citet{dav78} identified G206.9+2.3 as a distinct SNR from
  Monoceros based on optical observation. The SNR is very close to Monoceros
  and has a bright northwestern shell \citep{Gao11}. \citet{gra82} estimated a
  distance between 3--5\,kpc by the $\Sigma-D$ relation. Assuming the SNR--MC
  association, \citet{SuY2017} suggested a kinematic distance of 1.6\,kpc to
  the SNR. We found the first conspicuous extinction jump at $0.89 \pm 0.02$\,kpc.
  The second jump around 3.2\,kpc is probably a fake feature because 1)
  not all the bands show a jump at this distance; 2) tracers with high extinctions
  further than 1.0\,kpc are generally on the same level, with no apparent jump
  after 3.0\,kpc. Our estimate is much smaller than previous results.
  But we still classify G206.9+2.3 as Level A with the same reasons for G93.7--0.2.
  The low surface brightness of G206.9+2.3 and the early $\Sigma - D$ relation could 
  introduce large uncertainty to the estimation of \citet{gra82}. The difference between 
  our result and \citet{SuY2017} may be due to the uncertainty of the rotation 
  curve towards the Galactic anti-center.

  \item{G213.0--0.6} is an old, shell-type SNR with extremely low radio surface
  brightness \citep{Reich03}. Its association with the $\HII$ region S284
  suggested a distance of 2.4\,kpc \citep{SP12}. While \citet{SuY2017} argued
  that there is no correlation between S284 and the SNR with the CO and radio
  observations, they obtained a kinematic distance of about 1.0\,kpc. We infer
  an extinction jump at $1.09 \pm 0.29$\,kpc. The very scattered distribution of
  tracers make the other tiny jump around 2.0\,kpc unreliable.

\end{itemize}

\subsection{Level B} \label{appsubsec:levelB}

\begin{itemize}

  \item{G65.3+5.7} is a typical evolved shell-type SNR with very low surface
  brightness at $\lambda 6$\,cm and $\lambda 11$\,cm \citep{Xiao09}. The referred
  region is bigger than the faint outline of the SNR seen in the radio map
  (Figure \ref{fig:snrReg}). But the stars inside the outlines are too few,
  so we take all the sources in the reference region into account. Our model fitting
  yields two
  distance components at 1.51\,kpc and 3.61\,kpc respectively. Considering the
  latitude of G65.3+5.7, a distance of 3.61\,kpc means $\sim$360\,pc above the
  Galactic plane, much larger than the typical scale-height around 100\,pc of the
  dust disk \citep{DS01,Misiriotis2006}. Thus we suggest that the distance to
  G65.3+5.7 is $1.51 \pm 0.04$\,kpc, while the further extinction jump is caused
  by the incomplete tracers. \citet{boumis04} measured the expansion proper
  motion of the remnant's optical filamentary edge and  a global expansion
  velocity of 155\,$\rm km\,s^{-1}$ which implied a distance of $0.77 \pm 0.20$\,kpc.
  But this expansion velocity is much smaller than a few other measurements which
  suggest a value of up to 400\,$\rm km\,s^{-1}$. Even with the lower limit of the
  velocity of 200\,$\rm km\,s^{-1}$ given by \citet{Lozinskaia80},  the distance
  should be larger than 0.99\,kpc, in agreement with our estimation of 1.51\,kpc if
  the error is taken into account. This case serves as an example to illustrate the
  uncertainty of the distance derived from the proper motion and shock velocity since
  both parameters can be quite uncertain. This SNR is classified
  as Level B because there are too few tracers and they are not in the
  central part.

  \item{G74.0--8.5} (The Cygnus Loop). G74.0--8.5 is a well studied object in all
  observable bands, from X-ray \citep{koo16} through optical \citep{Blair05} to
  radio \citep{sun06}. \citet{Minkowski58} calculated a kinematic distance of
  770\,pc. \citet{Blair05} reported a distance of 540\,pc to the Cygnus Loop,
  by measuring the proper motion and shock velocity with the HST data. \citet{Fesen18}
  found three stars associated with the expanding shell of the Cygnus Loop
  remnant and determined its distance to be $735 \pm 25$\,pc from the {\it Gaia}
  distances of these stars. In the region of the Cygnus Loop, we found many
  sample stars with relatively low extinction extending to over 6\,kpc.
  Higher extinction stars start to appear from 1\,kpc and at any distance until
  5\,kpc. Limited by the number of the nearby sample stars, the model fitting
  yielded null result on the distance, while a distance upper limit can be set
  at 1\,kpc for the Cygnus Loop because the higher extinction stars start to
  appear from 1\,kpc and this is also consistent with all the other estimates.

  \item{G82.2+5.3} (W63). \citet{Mavromatakis04} made a multi-wavelength study
  of G82.2+5.3, and suggested a distance between 1.6 and 3.3\,kpc by the
  Sedov analysis. \citet{RG81} suggested a distance of 1.6\,kpc by measuring
  the expansion velocity of optical filaments. Our fitting results in a distance
  of $1.34 \pm 0.13$\,kpc to W63 that is in line with these two estimates,
  while much smaller than the result of \citet{shan18} that probed a
  distance of $3.2
  \pm 0.4$\,kpc using RC stars. As W63 has not been verified to be associated
  with MCs and it has a comparatively similar $\agra$ to $\asil$,
  this case is classified as Level B.

  \item{G108.2--0.6} is firstly identified as a faint and large shell-type SNR
  by \citet{Tian07}. There are many bright objects around this SNR, such as
  SNR G109.1--1.0 (southeast), the bright $\HII$ region Sh2--142 (southwest), and
  the MC Sh2--152 (south). Based on the $\HI$ observations,
  \citet{Tian07} suggested G108.2--0.6 is located in the Perseus arm, with
  a distance of $3.2 \pm 0.6$\,kpc. Our analysis suggests a much smaller distance at
  $1.02 \pm 0.01$\,kpc, which is well confined by many tracers with a strong
  extinction jump. The problem is the gap from $\sim$1.5\,kpc to about
  2.5\,kpc: there are no tracers with high extinctions in this range and the dwarfs
  are from the adjacent region rather than the SNR region. Furthermore, no associated
  MCs have been found. Thus the measured object could be a foreground
  cloud, and finally G108.2-0.6 is classified as Level B.

  \item{G113.0+0.2} with strong polarized emission was discovered by \citet{Kothes05},
  with the data of the Canadian Galactic Plane Survey. \citet{Kothes05}
  determined the distance of G113.0+0.2 as 3.1\,kpc through the observation of
  $\HI$ emission. Our result, $<$3.8\,kpc, is consistent with it.

  \item{G116.9+0.2} (CTB 1). \citet{Fich86} suggested that G116.9+0.2, G114.3+0.3,
  and G116.5+1.1 (for the latter two SNRs, see Section \ref{appsubsec:levelC}) belong
  to the same group within a large $\HI$ shell in the Perseus arm, and the distance
  to CTB 1 is estimated to be 4.2\,kpc. With the observational improvements and some
  additional constraints, \citet{Uyaniker04} moved these three SNRs from the
  Perseus arm to the Local arm by determining their distances as 0.7\,kpc for
  G114.3+0.3, 1.6\,kpc for G116.5+1.1 and G116.9+0.2. Our analysis based on one
  cloud component yields an estimate of distance at $4.3 \pm 0.2$\,kpc. If a
  two-component model is adopted, there will be a second component at around
  1\,kpc. It seems \citet{Fich86} and \citet{Uyaniker04} detected the close
  and far components of our two-component model respectively. The evidence to clarify
  which component is associated with the SNR is needed.

  \item{G127.1+0.5} is identified as a shell-type SNR by \citet{Caswell77} and
  \citet{Pauls77}. Several compact sources are near the center \citep{Kaplan04}
  and there is a fairly circular shell from north to northeast \citep{Sun07}.
  \citet{ZhouX2014} discovered a pre-existing molecular filament from the CO
  observation with which G127.1+0.5 is associated. This association implies a
  distance of $\sim$300\,pc. The possible association of the SNR with NGC 559
  \citep{RH07} and an $\HI$ bubble \citep{LT06} sets a distance of 1.3\,kpc
  \citep{ZhouX2014}. Meanwhile, \citet{LT06} suggested its distance to be 1.15\,kpc
  within an upper limit of 2.9\,kpc. Our sample stars set an upper limit at
  about 2.9\,kpc highly consistent with \citet{LT06}.

\end{itemize}

\subsection{Level C} \label{appsubsec:levelC}

\begin{itemize}

  \item{G65.1+0.6} is a faint shell-type SNR with strong southern emission
  \citep{landercker90}. Based on the associated $\HI$ observations, \citet{TL06}
  proposed a rather large distance of 9.2\,kpc. Our fitting shows a distance of
  1.33\,kpc. Since there are not enough tracers between 1.0 and 2.5\,kpc, the distance
  uncertainty may be up to 0.6\,kpc. It should be noticed that there are no
  tracers further than 7\,kpc due to the sensitivity limit of the spectroscopic
  surveys, consequently, the extinction jump at a distance like 9.2\,kpc would not
  show up  and the present detection of the jump at 1\,kpc might be caused by a local
  foreground MC.  The large discrepancy between our
  result and the kinematic estimate puts this case into Level C. Meanwhile there
  are no other reliable studies about its distance and the association with
  MCs.

  \item{G67.6+0.9} was discovered by \citet{Sabin13}. \citet{shan18} probed a
  distance of $3.2 \pm 0.4$\,kpc by RC stars. There are five tracers towards
  G67.6+0.9,
  and all of them exhibit much larger extinctions than ``$l165$''. So the distance
  upper limit is defined as the location of the nearest tracer, i.e. 2.6\,kpc
  although it is still much smaller than the distance derived from the RC
  stars.

  \item{G70.0--21.5 and G159.6+7.3}. Both are identified
  as SNR by the numerous optical filaments from the VTSS H$\alpha$ images
  \citep{Fesen10,fesen15}. But the available radio observations for G70.0--21.5
  (GB6S $\lambda 6$\,cm) and G159.6+7.3 (WENSS 325\,MHz) show no SNR-like structures.
  Our tracers neither exhibit any apparent extinction jump. The runs of reddening
  towards them after 1.0\,kpc are very flat, implying the deviation from the
  Galactic dust disk. Based on a shock velocity estimate, \citet{fesen15} derived a
  distance of 1--2\,kpc to G70.0--21.5. While for G159.6+7.3, no distance information
  is available. No reliable estimation can be made by our star sample.

  \item{G114.3+0.3 and G116.5+1.1}. \citet{RB81} discovered G114.3+0.3 and G116.5+1.1,
  and suggested their distance as 3.4\,kpc and 3.6--5.2\,kpc respectively based
  on the velocity of the possibly associated $\HI$ region. \citet{Fich86} suggested
  that these two SNRs, plus G116.9+0.2, belong to the same group within a large
  $\HI$ shell in the Perseus arm, with a distance of 4.2\,kpc. With the observational
  improvements and some additional constraints including the optical emission
  \citep{Fesen97}
  and the polarization horizon \citep{Uyaniker03}, \citet{Uyaniker04} moved these
  three SNRs from the Perseus arm to the Local arm after determining their distances
  as 0.7\,kpc for G114.3+0.3, 1.6\,kpc for G116.5+1.1 and G116.9+0.2. Our data set
  a distance upper limit of G114.3+0.3 to be 2.7\,kpc. Most of the tracers of
  G116.5+1.1
  are further than 3.0\,kpc, while the nearest one indicates a distance of 0.68\,kpc
  which is consistent with the distance derived from the 3D dust map (see Section
  \ref{subsubsec:distCompare}). So G116.5+1.1 is suggested to be located either at
  0.68\,kpc
  or within 3.0\,kpc. The estimates to both of these 2 SNRs are consistent
  with the results of \citet{Uyaniker04}. In Section \ref{appsubsec:levelB},
  G116.9+0.2 is determined as far as $4.3 \pm 0.2$\,kpc, consistent with
  \citet{RB81} while further than \citet{Uyaniker04}. Our results provide the
  possibility that these three SNRs are located in the Perseus arm. But because no
  foreground tracers are available, we can not exclude the possibility that G114.3+0.3
  and G116.5+1.1 may be located in the Local arm.

  \item{G119.5+10.2} (CTA 1), discovered by \citet{HR60}, is a composite SNR,
  with semi-circular shell and strong optical filaments \citep{SunXH11}. $\HI$
  observations are used to derive a distance of $1.4 \pm 0.3$\,kpc \citep{Pineault93}.
  Our tracers present a flat reddening profile with almost no increase of
  extinction to a distant zone. Thus no reliable estimate can be made.

  \item{G178.2--4.2}. \citet{Gao11b} identified this SNR, and studied its radio
  properties. No $\HI$ cavity is detected towards the sight line of G178.2--4.2.
  Neither the $\Sigma - D$ relation can yield a reasonable distance \citep{Gao11b}.
  High extinction tracers in our sample have a wide distribution in distance
  from 0.3\,kpc to 5\,kpc, but there is no apparent extinction jump along the
  distance. Consequently the distance to G178.2--4.2 cannot be well determined,
  and we suggest this SNR is located within 0.3\,kpc or further than 5\,kpc.

  \item{G179.0+2.6} was first identified as a thick shell-type SNR by \citet{FR86}.
  The SNR exhibits coincident optical emission consisting of diffuse filamentary
  features \citep{How2018}. The distance measurement based on the $\Sigma - D$
  relation is very uncertain in a wide range: 3.5\,kpc \citep{FR86},
  6.1\,kpc \citep{CB98}, 2.9\,kpc \citep{Guseinov03a}, 3.1\,kpc \citep{Pavlovic2014}.
  From our sample, a few tracers with high extinctions are further than 3\,kpc,
  but our model determines its distance at $0.92 \pm 0.04$\,kpc with a
  relatively small extinction jump (see Fig. \ref{fig:snrDist}).
  The tracers around 2\,kpc with $\rm E(\gps-\Ks) > 1.0$\,mag smoothen the increase
  of extinction from nearby to over 3\,kpc so that the jump at 0.92\,kpc becomes
  significant. As a result, the tiny extinction jump may only indicate
  a foreground MC.

  \item{G180.0--1.7} (S147) is one of the most famous evolved SNRs in the Milky
  Way, with beautiful filaments in optical bands. It was first identified as a SNR
  by \citet{Minkowski58}. A lot of studies on the distance to S147 have been
  done. Based on $\Sigma-D$ relation, the distances are $0.8-1.37$\,kpc
  \citep{Kundu80}, 0.9\,kpc \citep{CC76}, 1.06\,kpc \citep{Guseinov03a}, and $1.6
  \pm 0.3$\,kpc \citep{Sofue80}. The dispersion is mainly due to the differences
  in models and observational data. By measuring the distance to the pulsar located
  in S147, three consistent results are obtained: 1.2\,kpc \citep{Kramer03},
  1.47$^{+0.42}_{-0.27}$\,kpc \citep{Ng07}, and 1.3$^{+0.22}_{-0.16}$\,kpc
  \citep{Chatterjee09}. Based on the detection of the pre-supernova binary companion
  HD 37424, \citet{Din15} suggested a distance of 1.333$^{+0.103}_{-0.112}$\,kpc.
  Using multiband photometric data from the Xuyi Schmidt Telescope Photometric Survey
  of the Galactic Anticentre (XSTPS--GAC), 2MASS and {\it WISE}, \citet{chen17}
  investigated
  some dense region in S147, and obtained a distance of $1.22 \pm 0.21$\,kpc. All
  these derived distances are around 1.2\,kpc. Our model yields two clouds at
  the distances of $0.38 \pm 0.10$\,kpc and $2.51 \pm 0.78$\,kpc respectively, neither
  of which agrees with previous consensus at 1.2\,kpc. From Fig. \ref{fig:snrDist},
  it can be seen that the smaller distance suffers large uncertainty, and may trace
  a foreground cloud which is also reported by \citet{chen17}. The larger distance
  is probably caused by the Perseus arm. Nevertheless, many stars with high extinction
  ($E(\gps-\Ks)>1.0$\,mag) are widely distributed between $\sim$0.2$-$1.5\,kpc and
  located
  in the whole western part of S147. It seems that our extinction--distance model is
  not able to clearly peel S147 from the foreground cloud, which exposes the inability
  of
  our method to the old SNRs with weak extinction.

\end{itemize}

\bibliographystyle{aasjournal}
\bibliography{bibfile}


\begin{deluxetable*}{ccccccccccc}
\tablecaption{Fitting Coefficients of Intrinsic Color Indices\label{tab:coeICI} to
Eq.\ref{eq:IciFit}}
\tabletypesize{\scriptsize}

\tablehead{
      & $\feh$        & \multicolumn3c{[--1.00, --0.75]} & \multicolumn3c{(--0.75,
      --0.50]} & \multicolumn3c{(--0.50, --0.25]} \\
      & Coefficients  & $A_0$ & $A_1$ & $A_2$ & $A_0$ & $A_1$ & $A_2$ & $A_0$ & $A_1$ &
      $A_2$
}
\startdata
dwarf & $(\gps-\Gbp)_0$ & 3.3965 & 1642.0 & --0.1198 & 4.8156 & 1337.7 & --0.1111 &
18.336 & 948.82 & --0.0994 \\
      & $(\gps-\rps)_0$ & 6.8456 & 2387.4 & --0.3220 & 8.1132 & 2044.1 & --0.1799 &
      8.2938 & 2042.0 & --0.1735 \\
      & $(\gps-G)_0$    & 10.124 & 1722.8 & --0.0682 & 8.0118 & 1826.3 & --0.0766 &
      10.617 & 1629.8 & --0.0443 \\
      & $(\gps-\ips)_0$ & 9.8239 & 2481.3 & --0.5425 & 10.521 & 2284.1 & --0.4120 &
      10.533 & 2295.4 & --0.4007 \\
      & $(\gps-\Grp)_0$ & 10.762 & 2307.3 & --0.1377 & 11.827 & 2094.8 & --0.0008 &
      10.509 & 2289.4 & --0.0728 \\
      & $(\gps-J)_0$    & 15.609 & 2457.1 & --0.2175 & 15.855 & 2345.0 & --0.0679 &
      15.086 & 2445.1 & --0.1121 \\
      & $(\gps-H)_0$    & 18.059 & 2626.9 & --0.4107 & 17.568 & 2624.1 & --0.3438 &
      18.543 & 2486.5 & --0.1971 \\
      & $(\gps-\Ks)_0$  & 20.174 & 2469.5 & --0.3111 & 18.812 & 2563.9 & --0.3301 &
      19.969 & 2408.4 & --0.1413 \\
      & $(\gps-W_1)_0$  & 20.587 & 2452.3 & --0.2629 & 22.085 & 2261.9 & --0.0259 &
      20.018 & 2428.3 & --0.1260 \\
      & $(\gps-W_2)_0$  & 21.872 & 2354.5 & --0.2384 & 22.734 & 2204.9 & --0.0059 &
      19.799 & 2428.9 & --0.1472 \\
\hline
giant & $(\gps-\Gbp)_0$ & 13604. & 367.94 & --0.0337 & 37.086 & 784.62 & --0.0772 &
50.532 & 729.79 & --0.0521 \\
      & $(\gps-\rps)_0$ & 56.716 & 952.14 &  0.1888 & 17.020 & 1349.3 &  0.0955 & 45.508
      & 935.24 &  0.3492 \\
      & $(\gps-G)_0$    & 148.83 & 766.59 &  0.2257 & 47.262 & 875.52 &  0.1895 & 94.287
      & 809.30 &  0.3126 \\
      & $(\gps-\ips)_0$ & 110.00 & 893.97 &  0.3185 & 48.245 & 1090.4 &  0.2713 & 126.49
      & 817.69 &  0.5517 \\
      & $(\gps-\Grp)_0$ & 81.359 & 967.65 &  0.5911 & 54.792 & 1055.3 &  0.6085 & 93.782
      & 887.04 &  0.8110 \\
      & $(\gps-J)_0$    & 136.32 & 957.51 &  1.0640 & 33.534 & 1543.2 &  0.5248 & 85.929
      & 1030.1 &  1.2573 \\
      & $(\gps-H)_0$    & 151.16 & 994.32 &  1.2691 & 43.596 & 1507.3 &  0.7381 & 57.385
      & 1283.2 &  1.2327 \\
      & $(\gps-\Ks)_0$  & 183.70 & 952.47 &  1.3871 & 48.655 & 1472.0 &  0.7715 & 73.571
      & 1193.4 &  1.3819 \\
      & $(\gps-W_1)_0$  & 169.17 & 977.48 &  1.4094 & 45.457 & 1528.7 &  0.7448 & 111.16
      & 1044.6 &  1.6544 \\
      & $(\gps-W_2)_0$  & 168.39 & 969.26 &  1.3938 & 46.887 & 1475.8 &  0.8206 & 110.03
      & 1030.2 &  1.6545 \\
\hline
\hline
      & $\feh$          & \multicolumn3c{(--0.25, 0]}     & \multicolumn3c{(0, 0.25]}
      & \multicolumn3c{(0.25, 0.50]}   \\
      & Coefficients    & $A_0$ & $A_1$ & $A_2$          & $A_0$ & $A_1$ & $A_2$
      & $A_0$ & $A_1$ & $A_2$          \\
\hline
dwarf & $(\gps-\Gbp)_0$ & 20.971 & 931.52 & --0.1084 & 19.110 & 966.03 & --0.1204 &
17.835 & 949.16 & --0.1151 \\
      & $(\gps-\rps)_0$ & 7.4441 & 2213.3 & --0.2120 & 6.8197 & 2350.9 & --0.2362 &
      6.1199 & 2485.9 & --0.2428 \\
      & $(\gps-G)_0$    & 11.585 & 1605.0 & --0.0505 & 12.553 & 1574.5 & --0.0517 &
      8.4923 & 1803.1 & --0.0690 \\
      & $(\gps-\ips)_0$ & 9.8520 & 2446.6 & --0.4581 & 9.0829 & 2608.5 & --0.5086 &
      8.1627 & 2775.8 & --0.5330 \\
      & $(\gps-\Grp)_0$ & 9.8195 & 2443.9 & --0.1390 & 9.4192 & 2522.7 & --0.1604 &
      7.6727 & 2872.5 & --0.2283 \\
      & $(\gps-J)_0$    & 13.924 & 2653.2 & --0.2475 & 12.894 & 2842.6 & --0.3504 &
      11.733 & 2990.6 & --0.3670 \\
      & $(\gps-H)_0$    & 16.722 & 2753.7 & --0.4145 & 15.973 & 2852.3 & --0.4735 &
      14.814 & 2950.7 & --0.4802 \\
      & $(\gps-\Ks)_0$  & 18.026 & 2655.7 & --0.3501 & 16.813 & 2812.7 & --0.4547 &
      15.780 & 2859.3 & --0.4125 \\
      & $(\gps-W_1)_0$  & 18.272 & 2643.9 & --0.3003 & 16.936 & 2814.9 & --0.4122 &
      16.338 & 2813.1 & --0.3521 \\
      & $(\gps-W_2)_0$  & 17.817 & 2663.2 & --0.3203 & 16.283 & 2866.8 & --0.4480 &
      15.557 & 2879.8 & --0.3922 \\
\hline
giant & $(\gps-\Gbp)_0$ & 357.66 & 532.77 & --0.0141 & 15.649 & 1004.2 & --0.1105 &
1866.1 & 437.57 &  0.0190 \\
      & $(\gps-\rps)_0$ & 61.869 & 864.50 &  0.4104 & 96.074 & 779.73 &  0.4665 & 145.14
      & 706.00 &  0.5516 \\
      & $(\gps-G)_0$    & 248.56 & 652.00 &  0.4096 & 211.79 & 688.79 &  0.4016 & 297.70
      & 643.10 &  0.4728 \\
      & $(\gps-\ips)_0$ & 234.13 & 720.31 &  0.6417 & 558.47 & 615.34 &  0.7310 & 1329.6
      & 538.96 &  0.8152 \\
      & $(\gps-\Grp)_0$ & 198.68 & 750.36 &  0.9311 & 346.95 & 672.96 &  1.0061 & 1730.7
      & 519.88 &  1.1741 \\
      & $(\gps-J)_0$    & 255.12 & 782.73 &  1.5623 & 507.36 & 680.09 &  1.7068 & 1503.2
      & 564.41 &  1.9059 \\
      & $(\gps-H)_0$    & 172.74 & 903.35 &  1.7851 & 500.71 & 706.91 &  2.0963 & 882.81
      & 633.18 &  2.2639 \\
      & $(\gps-\Ks)_0$  & 209.43 & 873.48 &  1.8894 & 685.39 & 674.28 &  2.2283 & 1157.1
      & 610.78 &  2.3963 \\
      & $(\gps-W_1)_0$  & 291.84 & 810.53 &  2.0519 & 947.96 & 638.52 &  2.3613 & 959.31
      & 636.16 &  2.4246 \\
      & $(\gps-W_2)_0$  & 307.48 & 791.57 &  2.0310 & 1012.8 & 624.43 &  2.3192 & 774.61
      & 654.21 &  2.3124 \\
\enddata
\end{deluxetable*}

\begin{deluxetable}{lcclc}
\tablecaption{Distances of SNRs in Level A. \label{tab:snrDist1}}
\tablehead{
\colhead{SNR}   & \colhead{$D_{\rm thiswork}$} & \colhead{$D_{\rm literature}$} & Method & References \\
\colhead{Names} & \colhead{(kpc)}              & \colhead{(kpc)}                &        &
}
\startdata
G78.2+2.1   & 0.98            & 1.7--2.6, $1.5 \pm 0.4$ & associated object & 1, 2 \\
G89.0+4.7   & $2.3 \pm 0.3$   & $1.9^{+0.3}_{-0.2}$, $0.80\pm0.07$, 1.0--1.6 
                              & RCs\tablenotemark{a}, associated object, $\Sigma - D$ & 3-8 \\
G93.7--0.2  & $2.16 \pm 0.02$ & $1.5 \pm 0.2$           & kinematics & 9 \\
G94.0+1.0   & $2.53 \pm 1.08$ & 4.5                     & associated object & 10 \\
G109.1--1.0 & $2.79 \pm 0.04$ & 3.0, $4.0 \pm 0.8$, 6.0 & kinematics, RCs & 11-13 \\
G152.4--2.1 & $0.59 \pm 0.09$ & $1.1 \pm 0.1$, $\bf \le 1.0$ & kinematics, extinction & 14, \textbf{32} \\
G156.2+5.7  & $0.68 \pm 0.20$ & 0.3, 1--3, 1.3, 3       & associated object, kinematics, Sedov estimate & 15-18 \\
G160.9+2.6  & $0.54 \pm 0.10$ & $0.8 \pm 0.4$, 1.1,     & kinematics, Sedov estimate, & 19, 20, \\
            &                 & 1.3--1.8, $\ge 1.1$, \textbf{0.6}  
                              & $\Sigma - D$, associated object, extinction & 7, 21, \textbf{32} \\
G166.0+4.3  & $3.24 \pm 0.03$ & $4.5 \pm 1.5$           & kinematics & 22 \\
G182.4+4.3  & $1.05 \pm 0.24$ & $\ge 3$, $\sim$$\bf 1.1$ & Sedov estimate, extinction & 23, \textbf{32} \\
G189.1+3.0  & $1.80 \pm 0.05$ & 0.7--1.5, 1.9,          & kinematics, & 21, 24, \\
            &                 & 1.5, $\bf 1.73^{+0.13}_{-0.09}$ 
                              & $\Sigma - D$, associated object, extinction & 7, 8, 25, \textbf{32} \\
G190.9--2.2 & $1.03 \pm 0.01$ & $1.0 \pm 0.3$, $\bf 1.03^{+0.02}_{-0.08}$ & kinematics, extinction & 26, \textbf{32} \\
G205.5+0.5  & $1.13 \pm 0.01$ & $1.6 \pm 0.3$, 1.6, 1.5, $\bf 0.93^{+0.05}_{-0.08}$/$\bf 1.26^{+0.09}_{-0.10}$ 
                              & $\Sigma - D$, extinction & 27-29, \textbf{32} \\
G206.9+2.3  & $0.89 \pm 0.02$ & 3--5, 1.6                & $\Sigma - D$, kinematics & 28, 30 \\
G213.0--0.6 & $1.09 \pm 0.29$ & $\sim$1.0, 2.4, $\bf 1.15 \pm 0.08$ 
                              & kinematics, associated object, extinction & 30, 31, \textbf{32} \\
\enddata
\tablerefs{(1) \citet{Leahy2013}; (2) \citet{Landecker1980}; (3) \citet{shan18};
(4) \citet{Humphreys78}; (5) \citet{Willis73}; (6) \citet{CC76}; (7) \citet{CL79};
(8) \citet{Milne79}; (9) \citet{Uyaniker02}; (10) \citet{Foster05}; (11) \citet{Kothes02}; 
(12) \citet{Tian10}; (13) \citet{dur06}; (14) \citet{Foster2010}; (15) \citet{GF07}; 
(16) \citet{Reich92}; (17) \citet{Yamauchi99}; (18) \citet{Pfe91}; (19) \citet{LT07}; 
(20) \citet{Leahy87}; (21) \citet{Loz81}; (22) \citet{Landecker89}; (23) \citet{Kothes98}; 
(24) \citet{Ambrocio-Cruz2017}; (25) \citet{Fesen1984}; (26) \citet{Foster13}; (27) \citet{dav78}; 
(28) \citet{gra82}; (29) \citet{lea86}; (30) \citet{SuY2017}; (31) \citet{SP12}; 
\textbf{(32) \citet{YuB2019}}.} 
\tablenotetext{a}{red clump stars}
\end{deluxetable}

\begin{deluxetable}{lcclc}
\tablecaption{Distances of SNRs in Level B. \label{tab:snrDist2}}
\tablehead{
\colhead{SNR}   & \colhead{$D_{\rm thiswork}$} & \colhead{$D_{\rm literature}$} & Method & References \\
\colhead{Names} & \colhead{(kpc)}              & \colhead{(kpc)}                &        &
}
\startdata
G65.3+5.7   & $1.51 \pm 0.04$    & $0.77 \pm 0.20$      & proper motion & 1 \\
G74.0--8.5  & $<$1.0             & 0.77, $0.54^{+0.01}_{-0.008}$, $0.735 \pm 0.025$ 
                                 & kinematic, shock velocity, associated object & 2-4 \\
G82.2+5.3   & $1.34 \pm 0.13$    & 1.6--3.3, 1.6, $3.2 \pm 0.4$ & Sedov estimate, expansion velocity, RCs & 5-7 \\
G108.2--0.6 & $1.02 \pm 0.01$    & $3.2 \pm 0.6$        & kinematics & 8 \\
G113.0+0.2  & $<$3.8             & 3.1                  & kinematics & 9 \\
G116.9+0.2  & $4.3 \pm 0.2$      & 1.6, 4.2             & kinematics & 10, 11 \\
G127.1+0.5  & $<$2.9             & 0.3/1.3, 1.15        & associated object, kinematics & 12, 13 \\
\enddata
\tablerefs{(1) \citet{boumis04}; (2) \citet{Minkowski58}; (3) \citet{Blair05}; (4) \citet{Fesen18};
(5) \citet{Mavromatakis04}; (6) \citet{RG81};  (7) \citet{shan18}; (8) \citet{Tian07}; 
(9) \citet{Kothes05}; (10) \citet{Uyaniker04}; (11) \citet{Fich86}; (12) \citet{ZhouX2014};
(13) \citet{LT06}.}
\end{deluxetable}

\begin{deluxetable}{lcclc}
\tablecaption{Distances of SNRs in Level C. \label{tab:snrDist3}}
\tablehead{
\colhead{SNR}   & \colhead{$D_{\rm thiswork}$} & \colhead{$D_{\rm literature}$} & Method & References \\
\colhead{Names} & \colhead{(kpc)}              & \colhead{(kpc)}                &        &
}
\startdata
G65.1+0.6   & $1.33 \pm 0.60$    & 9.2                  & kinematics & 1 \\
G67.6+0.9   & $<$2.6             & $3.2 \pm 0.4$        & RCs        & 2 \\
G70.0--21.5 & --                 & 1--2                 & shock velocity & 3 \\
G114.3+0.3  & $<$2.7             & 0.7, 3.4, 4.2        & kinematics & 4-6 \\
G116.5+1.1  & $<$3.0/0.68        & 1.6, 3.6--5.2, 4.2   & kinematics & 4-6 \\
G119.5+10.2 & --                 & $1.4 \pm 0.3$        & kinematics & 7   \\
G159.6+7.3  & --                 &                      &            &     \\
G178.2--4.2 & $<$0.3/$>$5.0      &                      &            &     \\
G179.0+2.6  & $0.92 \pm 0.04$    & 3.5, 6.1, 2.9, 3.1   & $\Sigma - D$ & 8-11 \\
G180.0--1.7 & $0.38 \pm 0.10$    & 1.06, 0.9, 0.8--1.37, $1.6 \pm 0.3$, & $\Sigma - D$, & 10, 12, 13, 14, \\
            &                    & 1.2, $1.47^{+0.42}_{-0.27}$, $1.3^{+0.22}_{-0.16}$,  & pulsar distance, & 15-17, \\
            &                    & 1.333$^{+0.103}_{-0.112}$, $1.22 \pm 0.21$ & pre-companion, extinction & 18, 19 \\
\enddata
\tablerefs{(1) \citet{TL06}; (2) \citet{shan18}; (3) \citet{fesen15}; (4) \citet{Uyaniker04};
(5) \citet{RB81}; (6) \citet{Fich86}; (7) \citet{Pineault93}; (8) \citet{FR86}; (9) \citet{CB98};
(10) \citet{Guseinov03a}; (11) \citet{Pavlovic2014}; (12) \citet{CC76}; (13) \citet{Kundu80};
(14) \citet{Sofue80}; (15) \citet{Kramer03}; (16) \citet{Ng07}; (17) \citet{Chatterjee09};
(18) \citet{Din15}; (19) \citet{chen17};}
\end{deluxetable}

\begin{deluxetable}{lccccccc}
\tablecaption{The fitting results to the extinction curves of 32 SNRs, as well as
              ``$l165$'' and the Rosette Nebula. \label{tab:ext-fitting}}
\tablehead{
\colhead{SNR} & \colhead{Level}\tablenotemark{a} & \colhead{$\Rv$\,[CCM89]} &
\colhead{$\Rv$\,[dust model]} & $\alpha_{\rm gra}$ &
\colhead{$\alpha_{\rm sil}$} & \colhead{$\rm \agra$\tablenotemark{b}\,$(\micron)$} &
\colhead{$\rm \asil$\tablenotemark{c}\,$(\micron)$}
}
\startdata
G65.1+0.6  & C & $3.75^{+0.08}_{-0.09}$ & 3.99 & 3.15 & 2.93 & 0.0093 & 0.0101 \\
G65.3+5.7  & B & $4.55 \pm 0.26$        & 4.97 & 2.85 & 2.08 & 0.0105 & 0.0185 \\
G67.6+0.9  & C & $2.99^{+0.06}_{-0.07}$ & 3.27 & 3.95 & 2.16 & 0.0076 & 0.0170 \\ 
G70.0--21.5 & C & $2.30^{+0.03}_{-0.02}$ & 2.44 & 4.43 & 3.30 & 0.0071 & 0.0088 \\ 
G74.0--8.5  & B & $2.79 \pm 0.23$        & --    & --    & --    & --      & --      \\
G78.2+2.1  & A & $4.03^{+0.40}_{-0.38}$ & 4.47 & 3.29 & 1.69 & 0.0088 & 0.0283 \\ 
G82.2+5.3  & B & $3.41 \pm 0.25$        & 3.61 & 3.35 & 3.18 & 0.0087 & 0.0091 \\ 
G89.0+4.7  & A & $4.29^{+0.16}_{-0.15}$ & 4.86 & 3.87 & -0.82 & 0.0077 & 0.1615 \\ 
G93.7--0.2  & A & $3.99 \pm 0.07$        & 4.50 & 3.65 & 0.64 & 0.0080 & 0.0869 \\ 
G94.0+1.0  & A & $3.81^{+0.44}_{-0.42}$ & 4.22 & 3.36 & 1.90 & 0.0086 & 0.0223 \\ 
G108.2--0.6 & B & $6.44^{+0.41}_{-0.40}$ & 6.74 & 1.78 & 1.90 & 0.0253 & 0.0222 \\ 
G109.1--1.0 & A & $3.44^{+0.42}_{-0.38}$ & --    & --    & --    & --      & --      \\
G113.0+0.2 & B & $3.40 \pm 0.15$        & 3.78 & 3.83 & 1.45 & 0.0077 & 0.0376 \\ 
G114.3+0.3 & C & $3.48^{+0.07}_{-0.06}$ & 3.82 & 3.50 & 2.19 & 0.0083 & 0.0166 \\ 
G116.5+1.1 & C & $3.36 \pm 0.08$        & 3.73 & 3.80 & 1.62 & 0.0078 & 0.0308 \\ 
G116.9+0.2 & B & $3.02^{+0.27}_{-0.25}$ & --    & --    & --    & --      & --      \\
G119.5+10.2& C & $2.71^{+0.44}_{-0.43}$ & --    & --    & --    & --      & --      \\
G127.1+0.5 & B & $3.36^{+0.11}_{-0.10}$ & 3.73 & 3.82 & 1.56 & 0.0077 & 0.0329 \\ 
G152.4--2.1 & A & $2.52 \pm 0.23$        & 2.62 & 4.05 & 3.81 & 0.0074 & 0.0078 \\ 
G156.2+5.7 & A & $5.08^{+0.31}_{-0.32}$ & 5.79 & 3.21 & 0.13 & 0.0091 & 0.1204 \\
G159.6+7.3 & C & $2.81 \pm 0.01$        & 2.97 & 3.78 & 3.37 & 0.0078 & 0.0086 \\ 
G160.9+2.6 & A & $2.86 \pm 0.15$        & 2.92 & 3.88 & 4.36 & 0.0077 & 0.0071 \\ 
G166.0+4.3 & A & $6.61^{+0.54}_{-0.51}$ & --    & --    & --    & --      & --      \\
G178.2--4.2 & C & $3.11^{+0.07}_{-0.08}$ & 3.33 & 3.59 & 2.95 & 0.0081 & 0.0100 \\ 
G179.0+2.6 & C & $7.71^{+0.72}_{-0.67}$ & --    & --    & --    & --      & --      \\
G180.0--1.7 & C & $3.28 \pm 0.29$        & 3.33 & 3.84 & 5.39 & 0.0077 & 0.0065 \\ 
G182.4+4.3 & A & $4.78^{+0.35}_{-0.34}$ & 5.33 & 2.94 & 1.44 & 0.0101 & 0.0378 \\ 
G189.1+3.0 & A & $3.67^{+0.17}_{-0.16}$ & 4.05 & 3.42 & 2.01 & 0.0085 & 0.0198 \\ 
G190.9--2.2 & A & $2.79 \pm 0.06$        & 3.00 & 3.91 & 2.82 & 0.0076 & 0.0107 \\ 
G205.5+0.5 & A & $4.01 \pm 0.16$        & 4.50 & 3.52 & 1.01 & 0.0083 & 0.0618 \\ 
G206.9+2.3 & A & $2.81^{+0.20}_{-0.19}$ & 3.02 & 3.86 & 2.92 & 0.0077 & 0.0102 \\ 
G213.0--0.6 & A & $5.98^{+0.46}_{-0.45}$ & --    & --    & --    & --      & --      \\
Rosette    & -- & $3.46 \pm 0.04$        & 3.57 & 3.38 & 3.76 & 0.0086 & 0.0078 \\ 
``$l165$'' & -- & $3.05^{+0.02}_{-0.03}$ & 3.23 & 3.59 & 3.34 & 0.0081 & 0.0087 \\ 
\enddata
\tablenotetext{a}{The quality of the distance measurement.}
\tablenotetext{b}{The average size of the graphite grains.}
\tablenotetext{c}{The average size of the silicate grains.}
\end{deluxetable}

\newpage


\begin{figure}[ht]
  \centering
  \includegraphics[width=16.8cm]{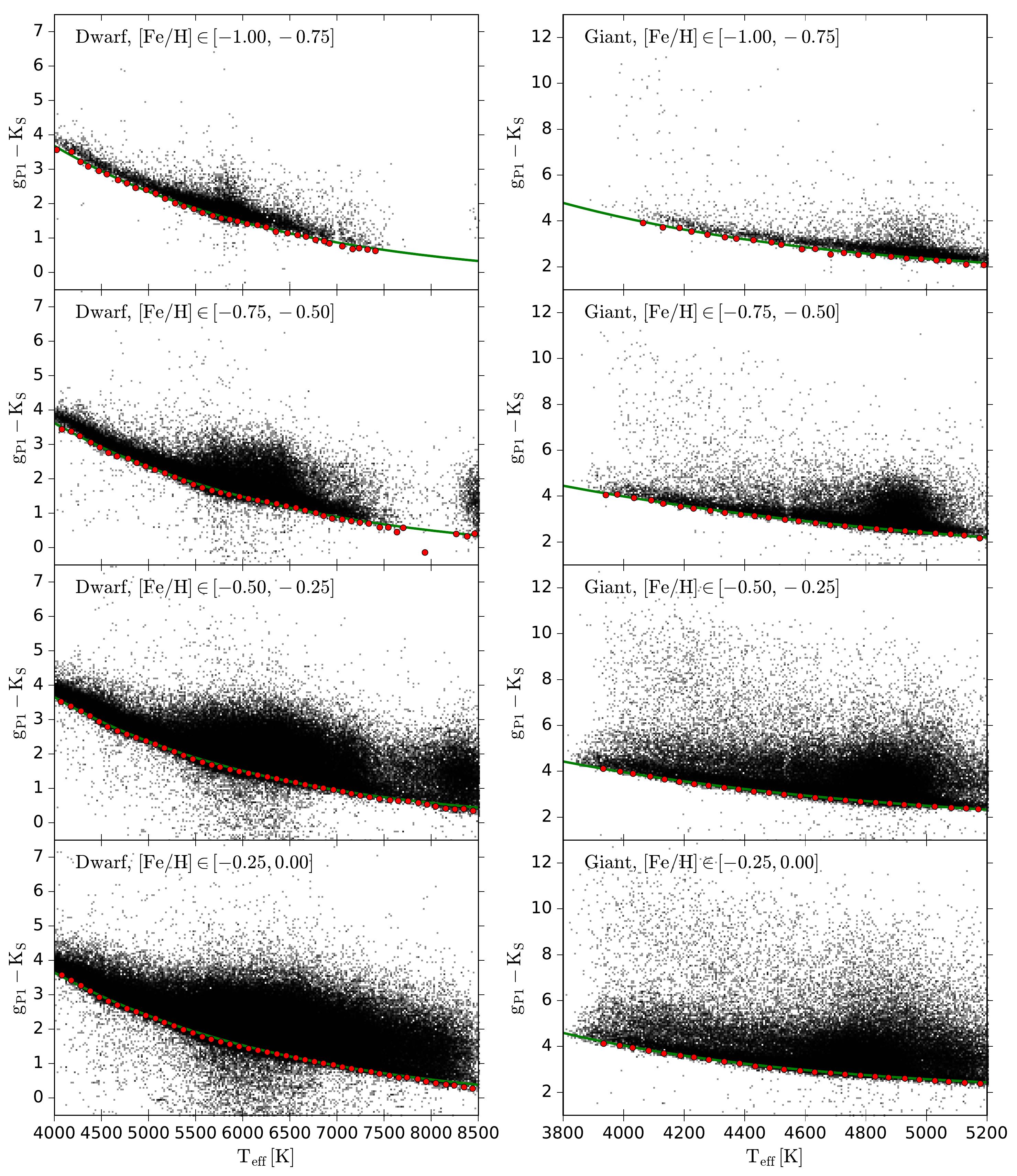}
  \caption{Determination of the intrinsic color index $(\gps-\Ks)_0$ with $\Teff$
  for both dwarfs (left panels) and giants (right panels) in different $\feh$ groups.
  The red dots denote the intrinsic colors derived in each $\Teff$ bin, and the
  green lines represent the fitting curves. The luminosity class and $\feh$ group
  are indicated in the top left corner on each panel. The last two panels compare
   $(\gps-\Ks)_0$ derived for different $\feh$ group for both dwarf
  (left panel) and giant (right panel).}
  \label{fig:ici}
\end{figure}\addtocounter{figure}{-1}

\begin{figure}[ht]
  \centering
  \includegraphics[width=16.8cm]{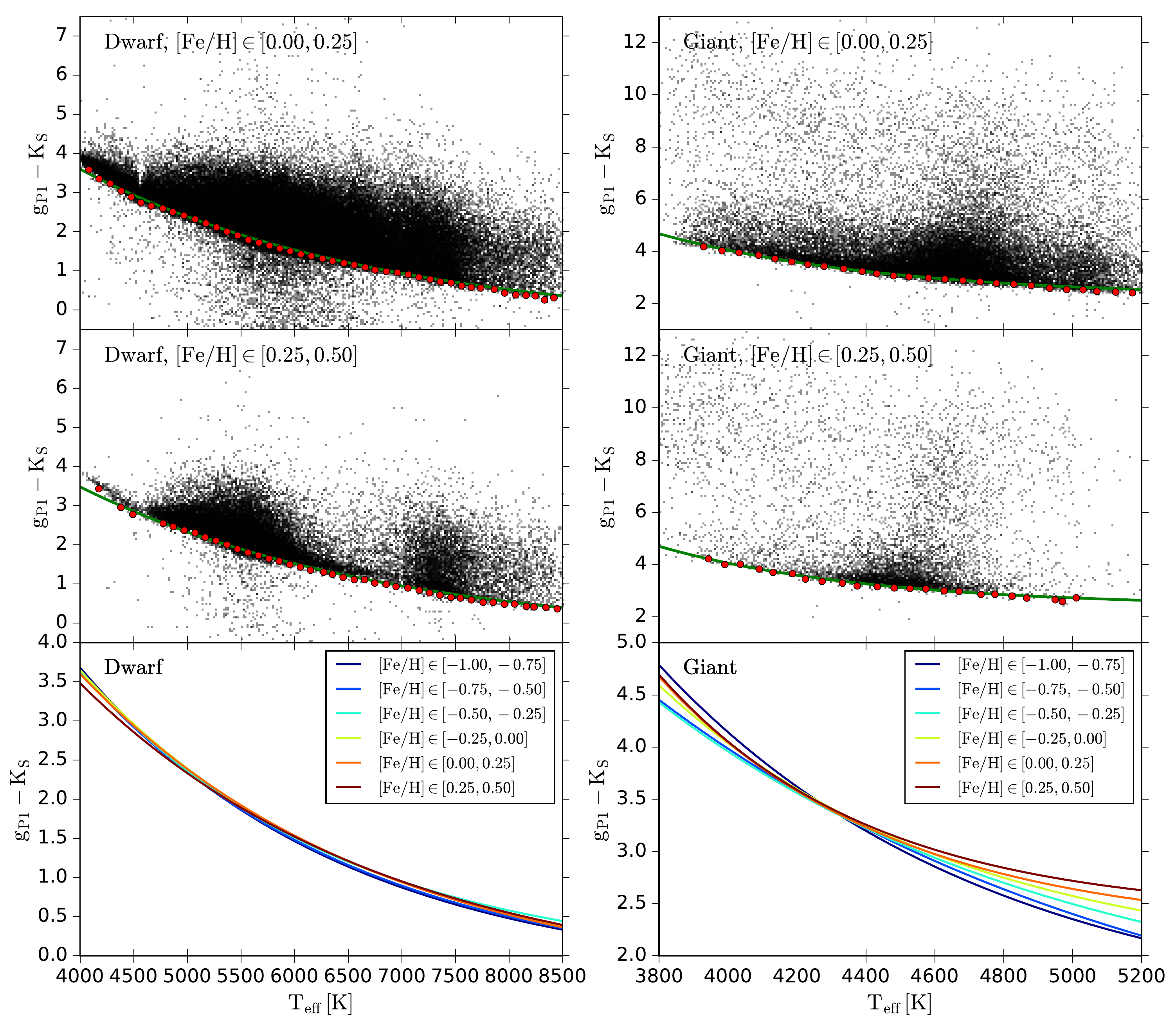}
  \caption{--continued.}
\end{figure}

\begin{figure}[ht]
  \centering
  \includegraphics[width=16.8cm]{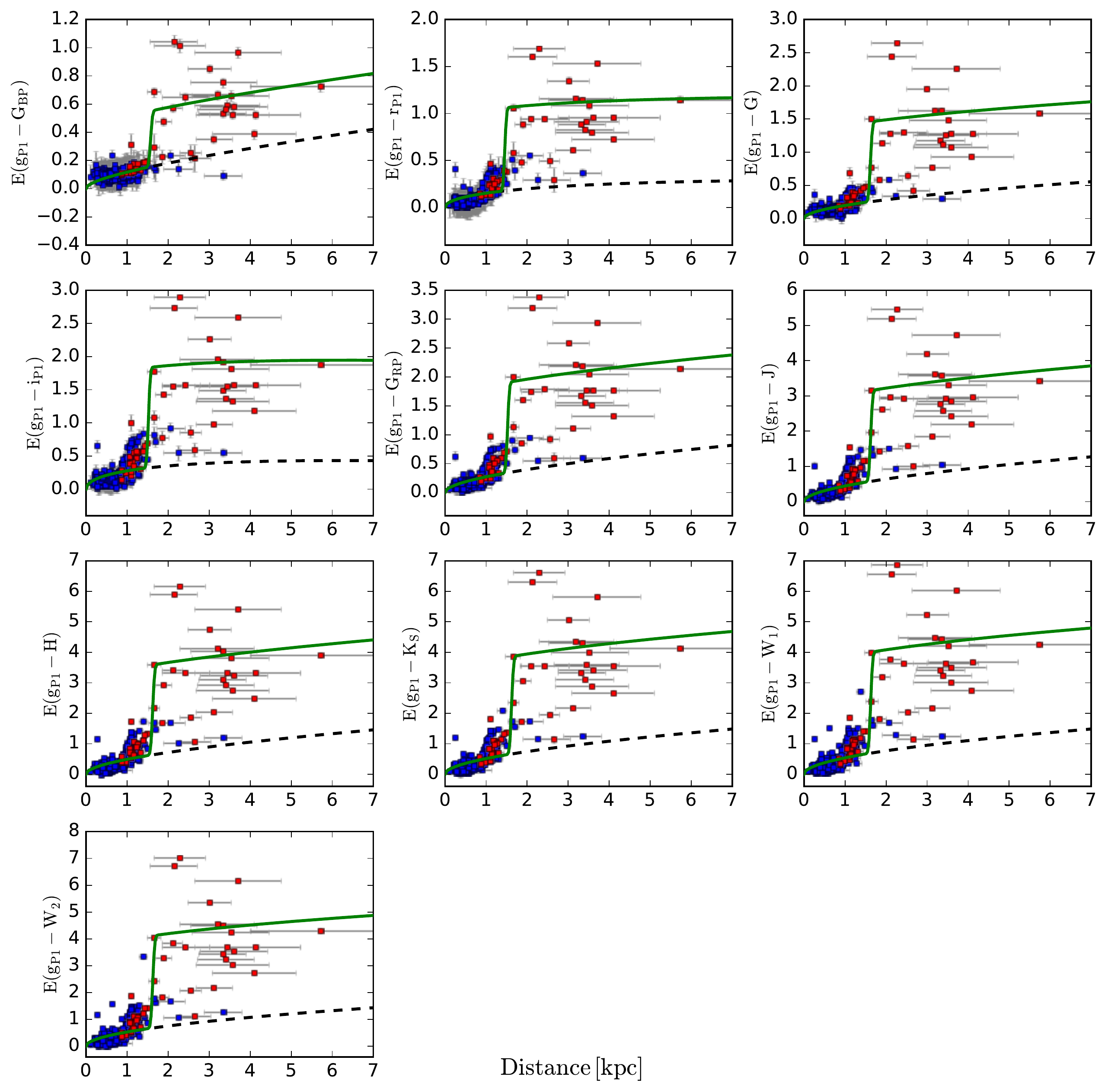}
  \caption{The fitting result of the extinction--distance model (Eq.\ref{eq:total-ext},
  \ref{eq:ext-dist}
  and \ref{eq:ism-profile}) to the Rosette Nebula in the 10 color excesses. The blue and
  red squares are
  dwarf and giant stars respectively. The gray lines stand for errorbars. The
  green lines are best fitting curves. The black dashed line is the ISM component
  derived from the extinction--distance model.}
  \label{fig:Rosette}
\end{figure}

\begin{figure}[ht]
   \centering
   \includegraphics[width=16.8cm]{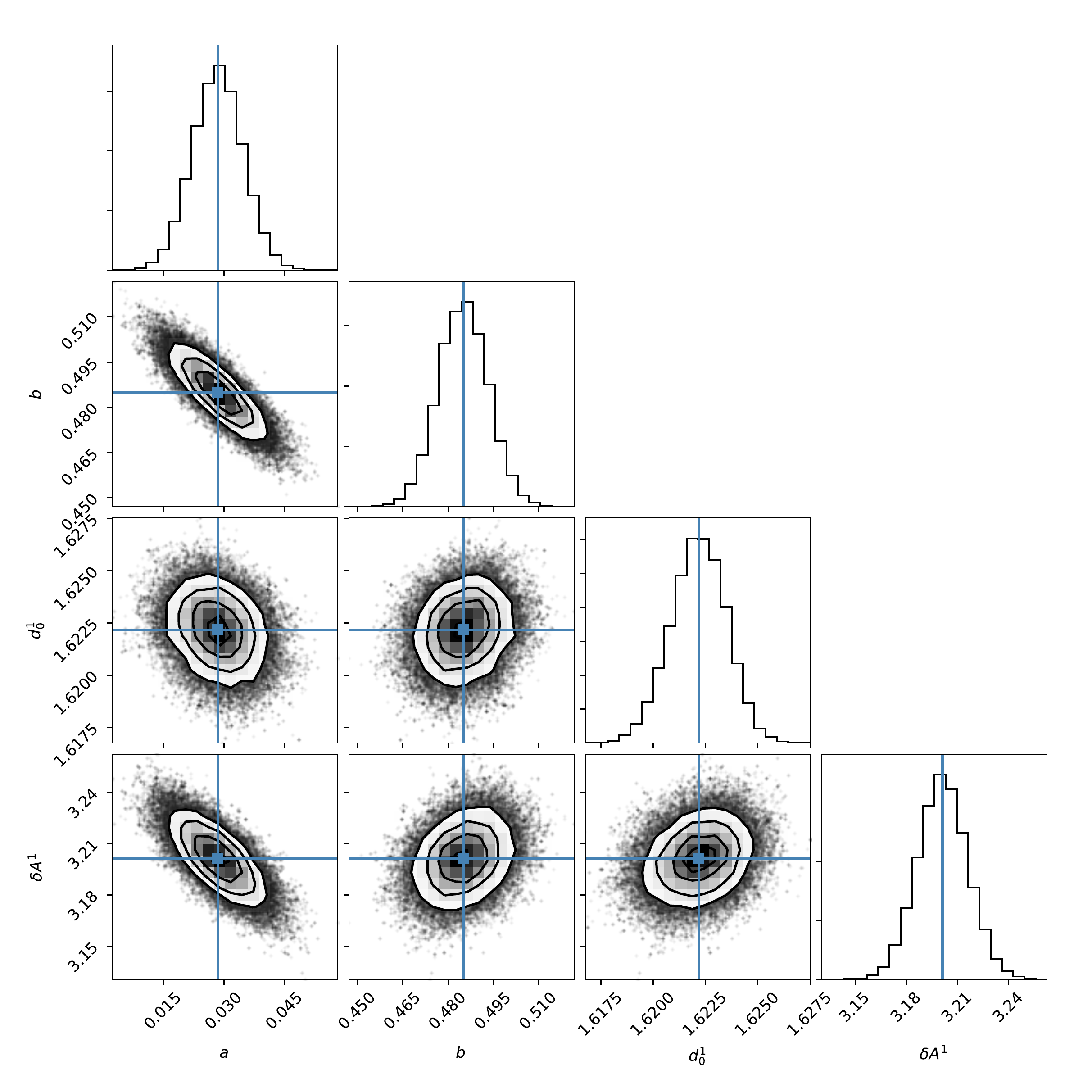}
   \caption{The corner plot of the fitting to the Rosette Nebula in $E(\gps-\Ks)$.
   The histograms show the distributions of the parameters. The contours present
   the covariances between each of them. The blue squares and lines indicate the
   best-fit values of the parameters.}
   \label{fig:corner}
\end{figure}

\begin{figure}[ht]
   \centering
   \includegraphics[width=16.8cm]{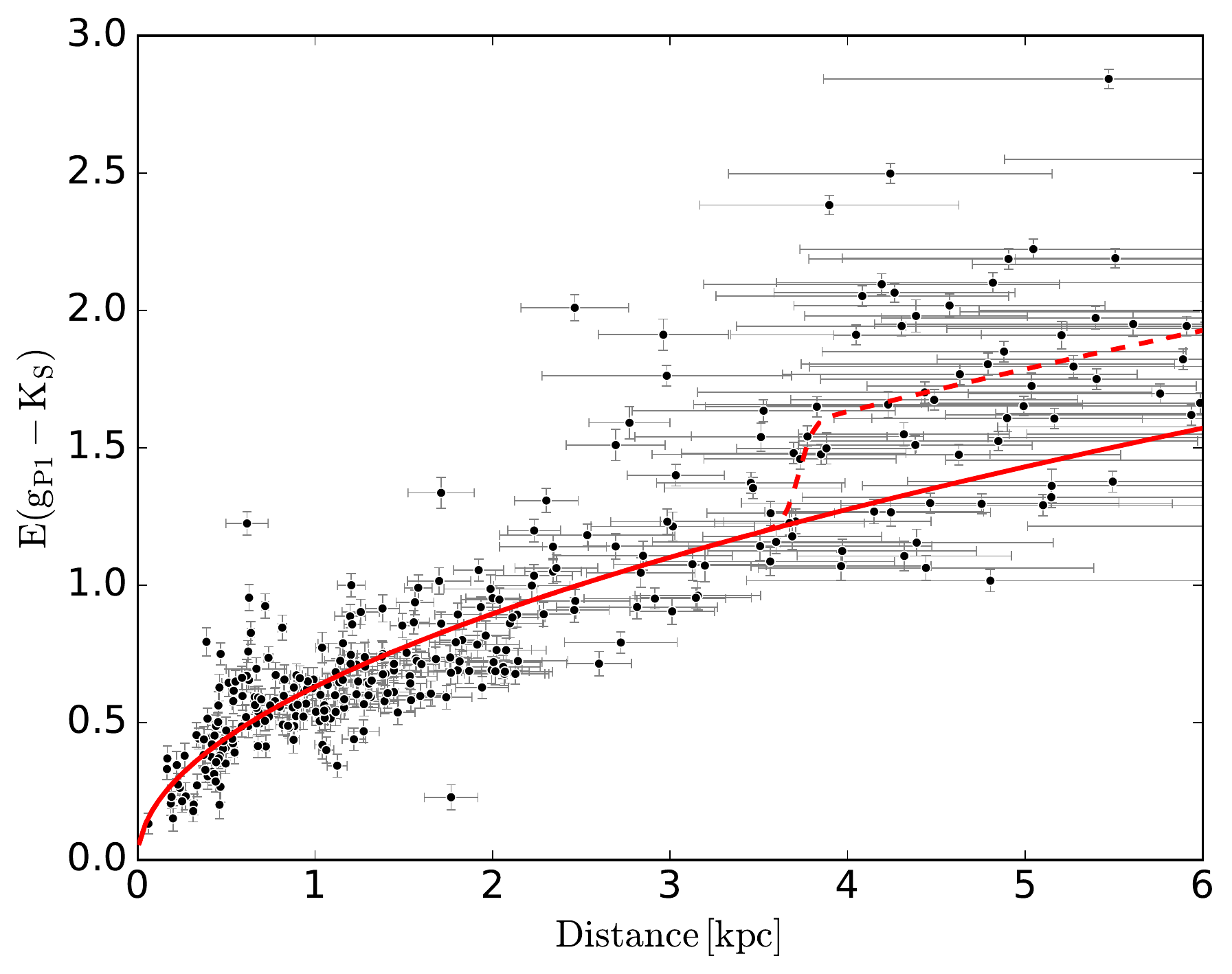}
   \caption{The run of reddening towards ``$l165$'', the common reference region,
   in $E(\gps-K_S)$. The black dots with errorbars are the sample stars.
   The solid red line is the best fitted reddening profile, and the dashed red
   line is used to describe the extinction jump caused by the Perseus arm.}
   \label{fig:CI}
\end{figure}

\begin{figure}[ht]
   \centering
   \includegraphics[width=16.8cm]{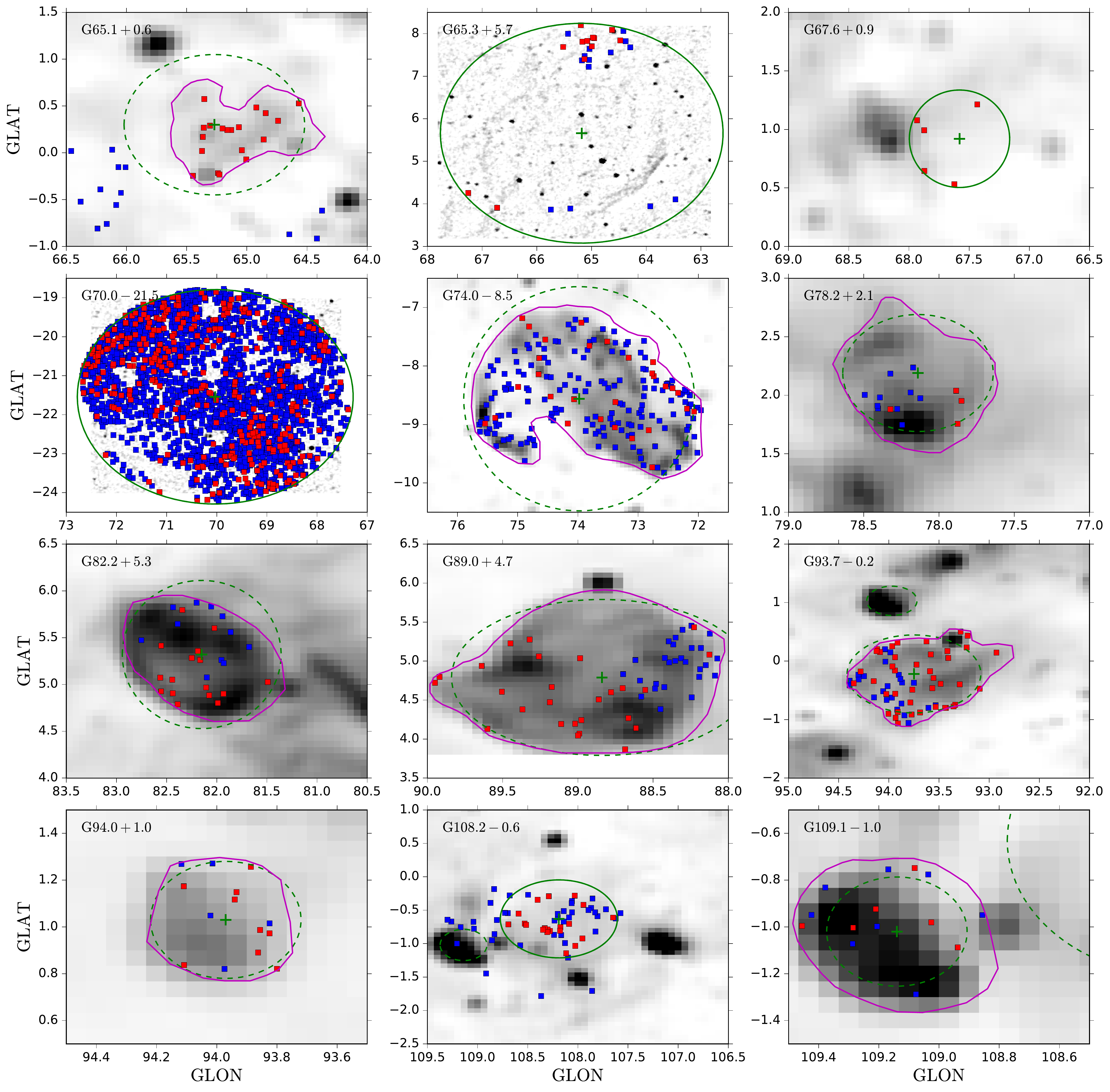}
   \caption{The selected stars and the SNR regions. The background gray image is the
   radio map. The green dashed circles represent the
   reference regions of SNRs from \citet{Green19}, with green crosses indicating
   the centers. If the referred regions are used, they will be in solid lines.
   The magenta solid lines are manually defined regions, which follow some contour
   lines enclosing the SNRs. The blue and red squares denote the dwarfs and
   giants, respectively. All the sub panels are in Galactic coordinates.}
   \label{fig:snrReg}
\end{figure}\addtocounter{figure}{-1}

\begin{figure}[ht]
   \centering
   \includegraphics[width=16.8cm]{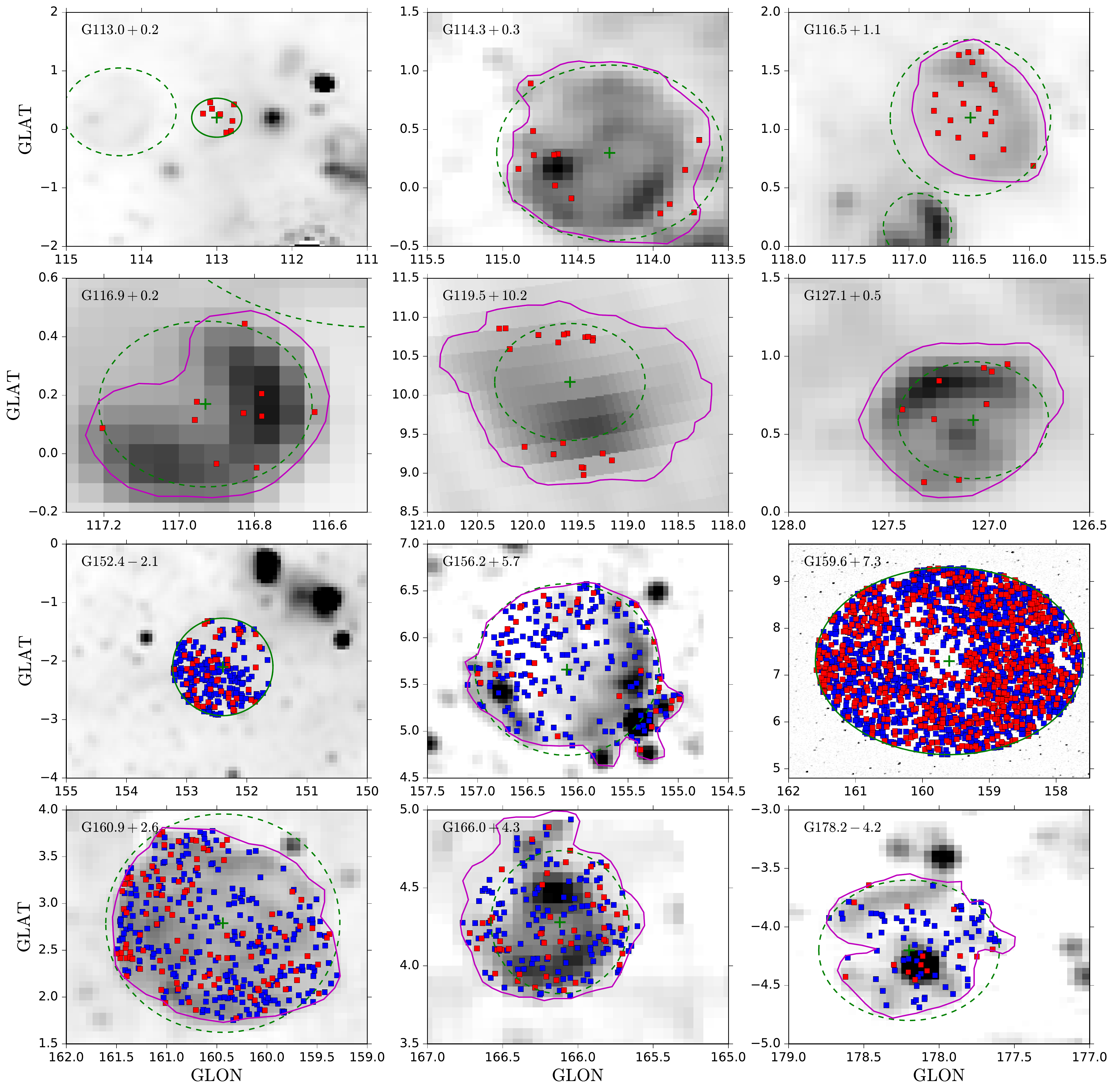}
   \caption{--continued.}
\end{figure}\addtocounter{figure}{-1}

\begin{figure}[ht]
   \centering
   \includegraphics[width=16.8cm]{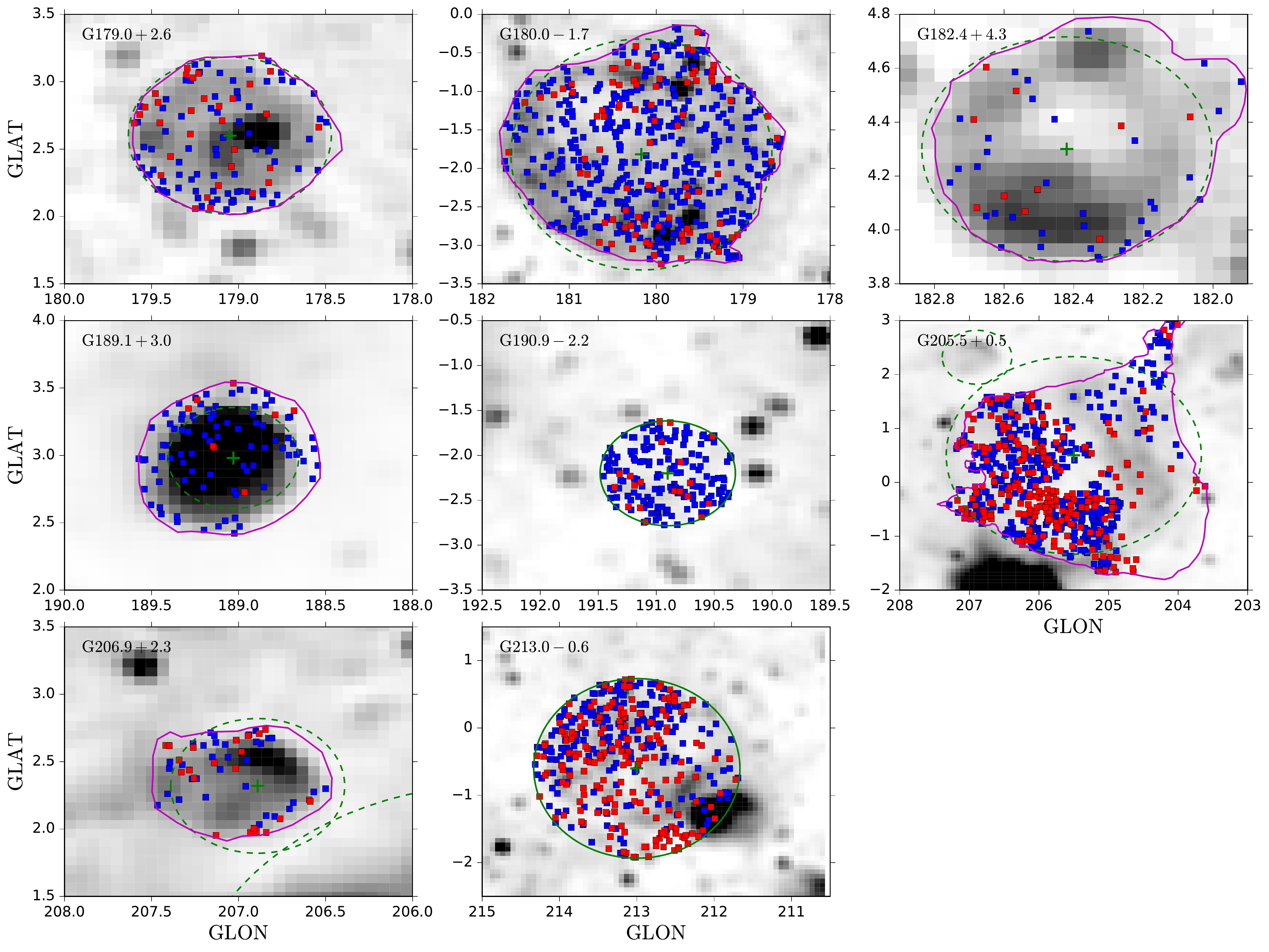}
   \caption{--continued.}
\end{figure}

\begin{figure}[ht]
   \centering
   \includegraphics[width=16.8cm]{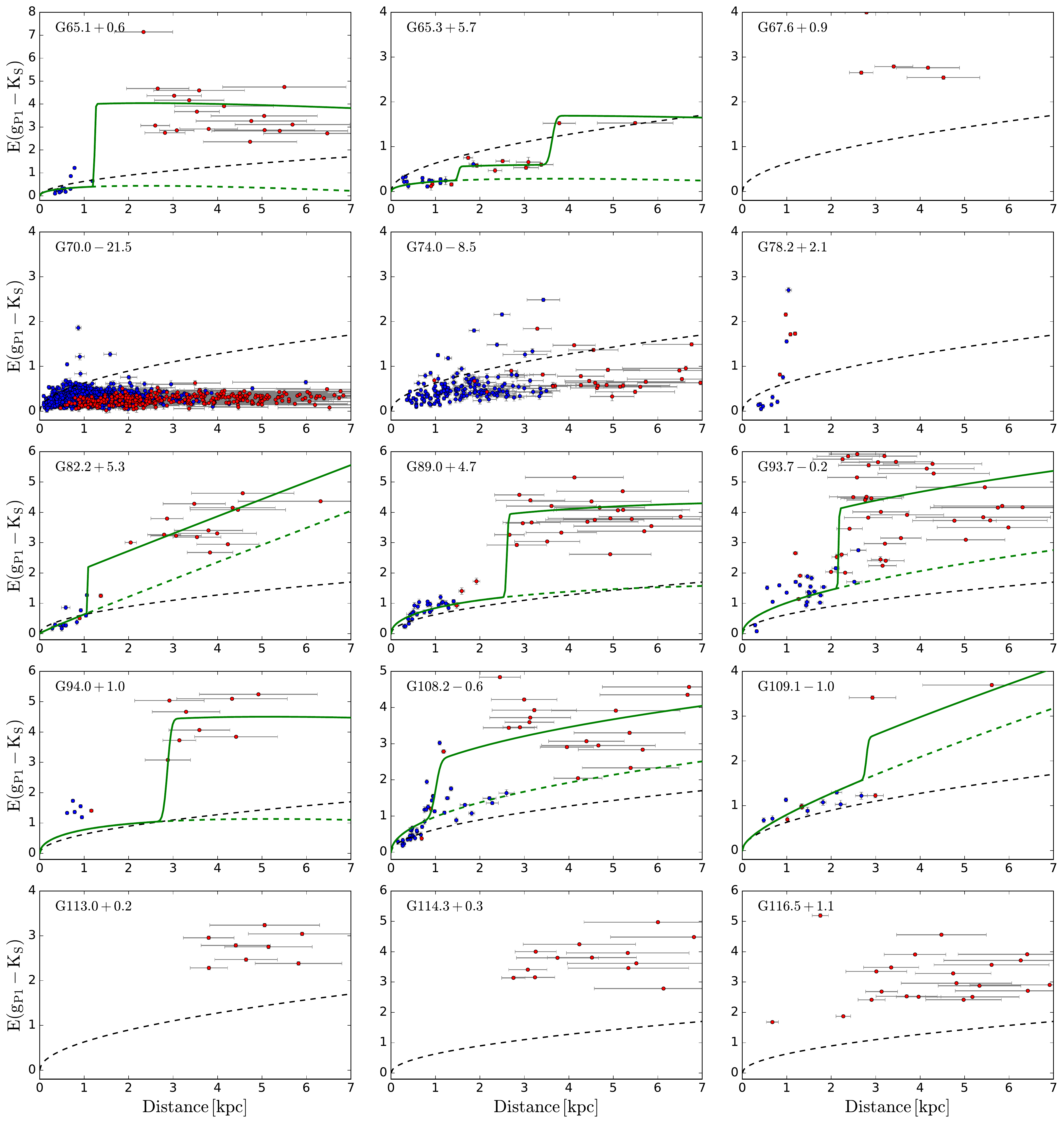}
   \caption{Color excess values $E(\gps-\Ks)$ versus distances (in
   kpc) in the selected regions for all the 32 SNRs. The blue and red dots represent
   dwarf and giant stars, respectively. The black dashed line is the derived
   common ISM reddening profile from the $l165$ region. The green solid lines are the
   best-fitting reddening
   profiles based on the sample stars. The green dashed lines decodes the ISM
   contribution
   derived from the extinction--distance model.}
   \label{fig:snrDist}
\end{figure}\addtocounter{figure}{-1}

\begin{figure}[ht]
   \centering
   \includegraphics[width=16.8cm]{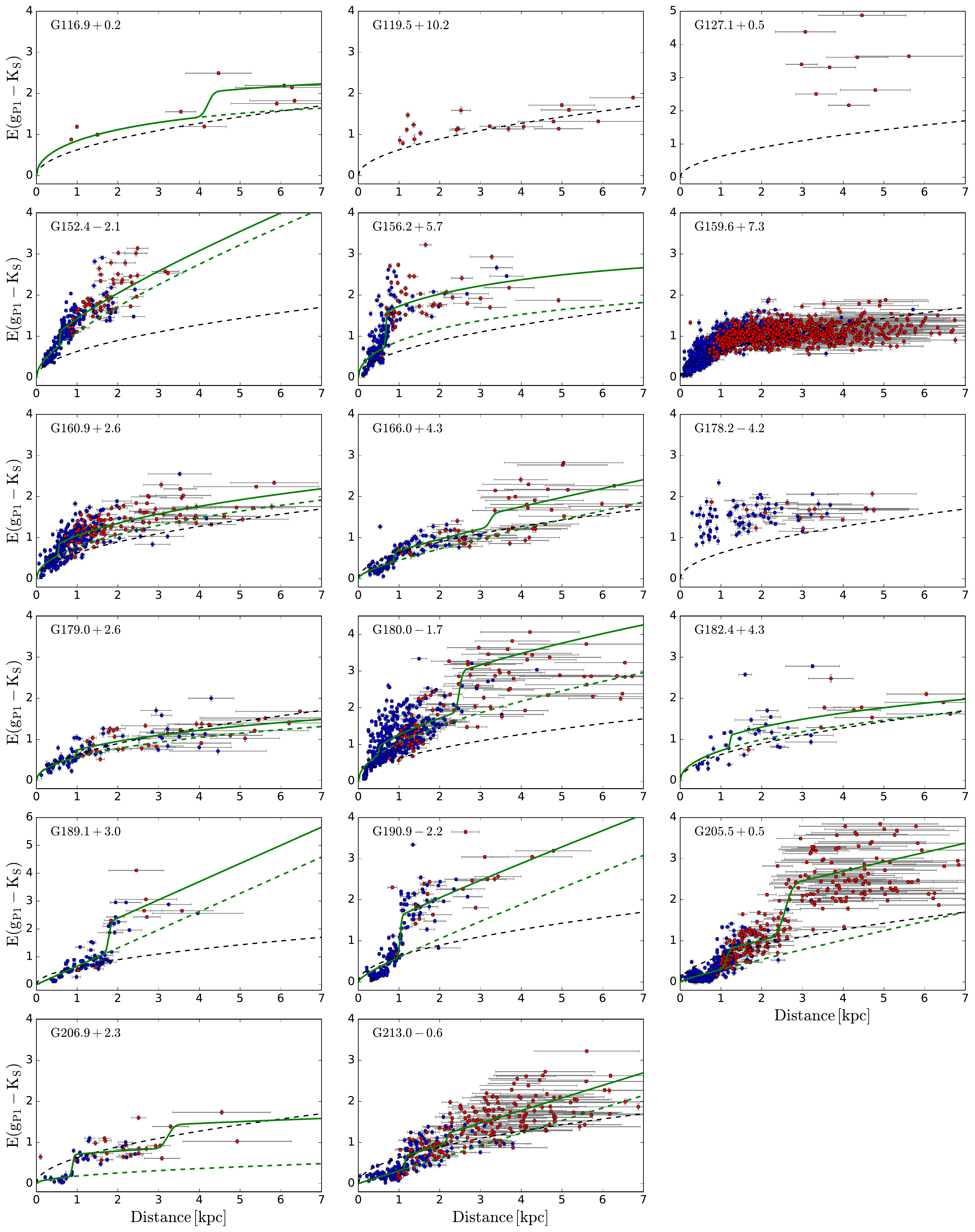}
   \caption{--continued.}
\end{figure}

\begin{figure}[ht]
   \centering
   \includegraphics[width=16.8cm]{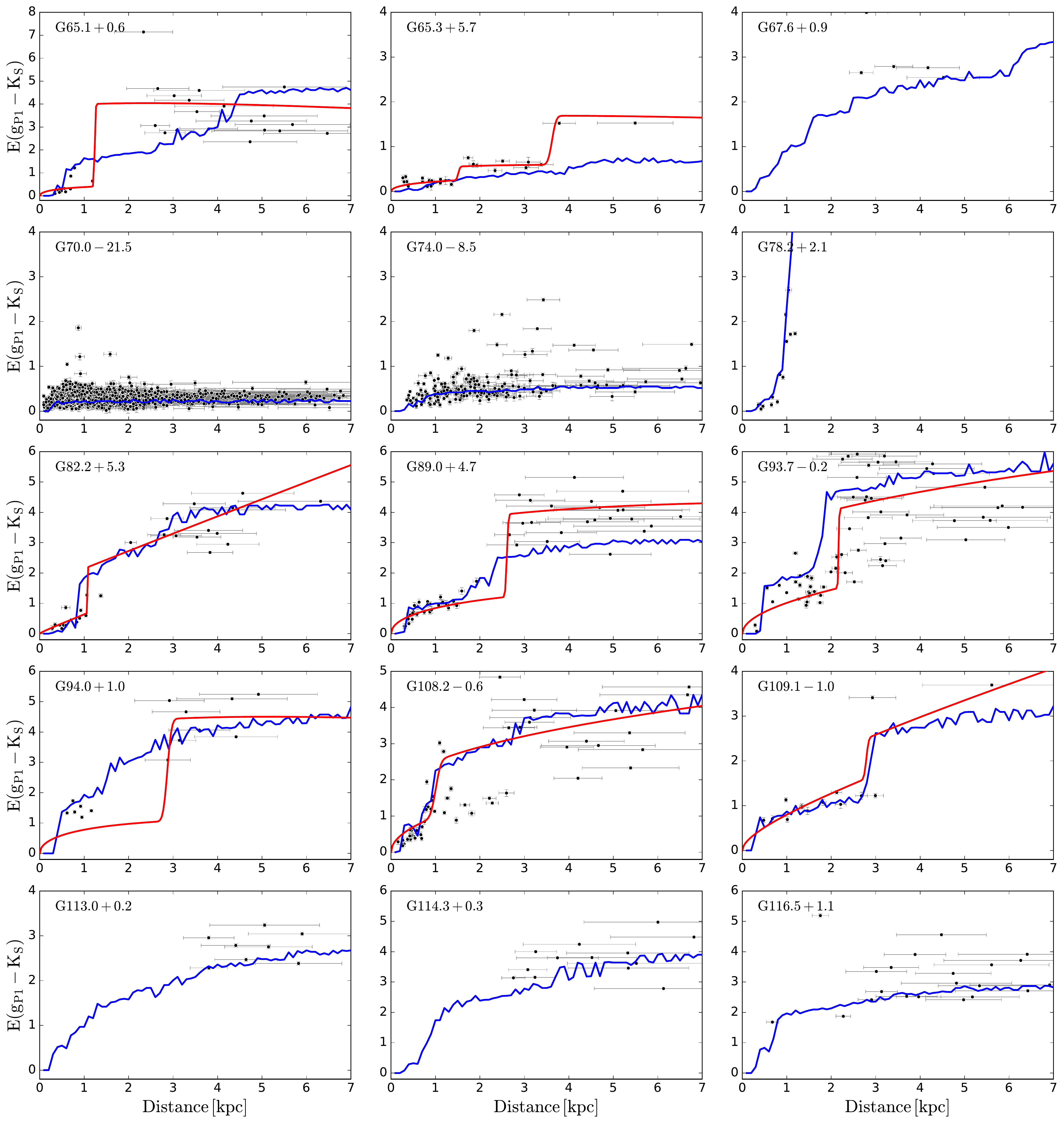}
   \caption{The comparison between our results (black dots and red lines) and
   the reddening profiles (blue lines) retrieved from the dust map  by
   \citet{Green2019}. The black dots and red lines are the same as the blue/red
   dots and green lines in Figure \ref{fig:snrDist}, respectively.}
   \label{fig:comGreen}
\end{figure}\addtocounter{figure}{-1}

\begin{figure}[ht]
   \centering
   \includegraphics[width=16.8cm]{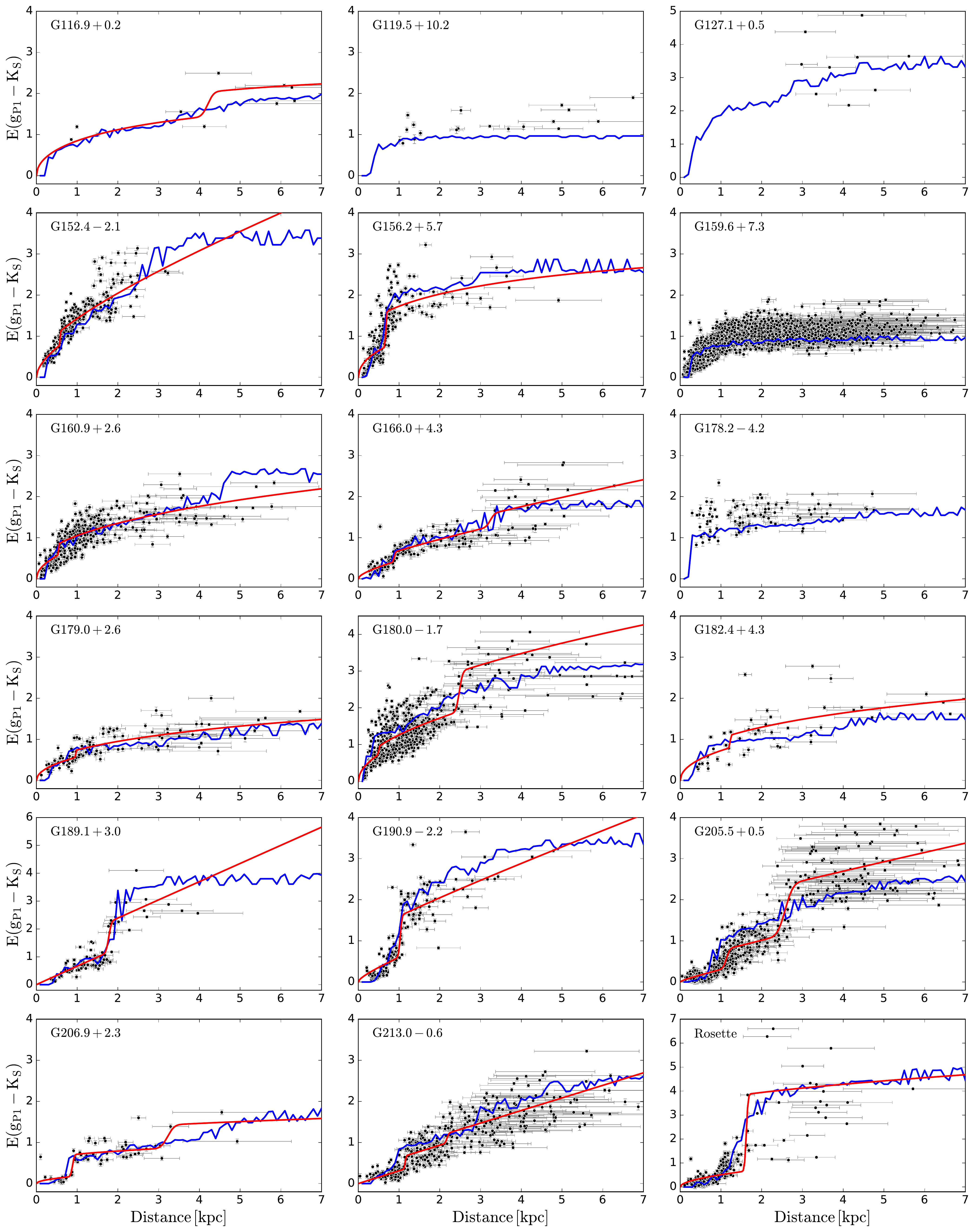}
   \caption{--continued.}
\end{figure}

\begin{figure}[ht]
   \centering
   \includegraphics[width=16.8cm]{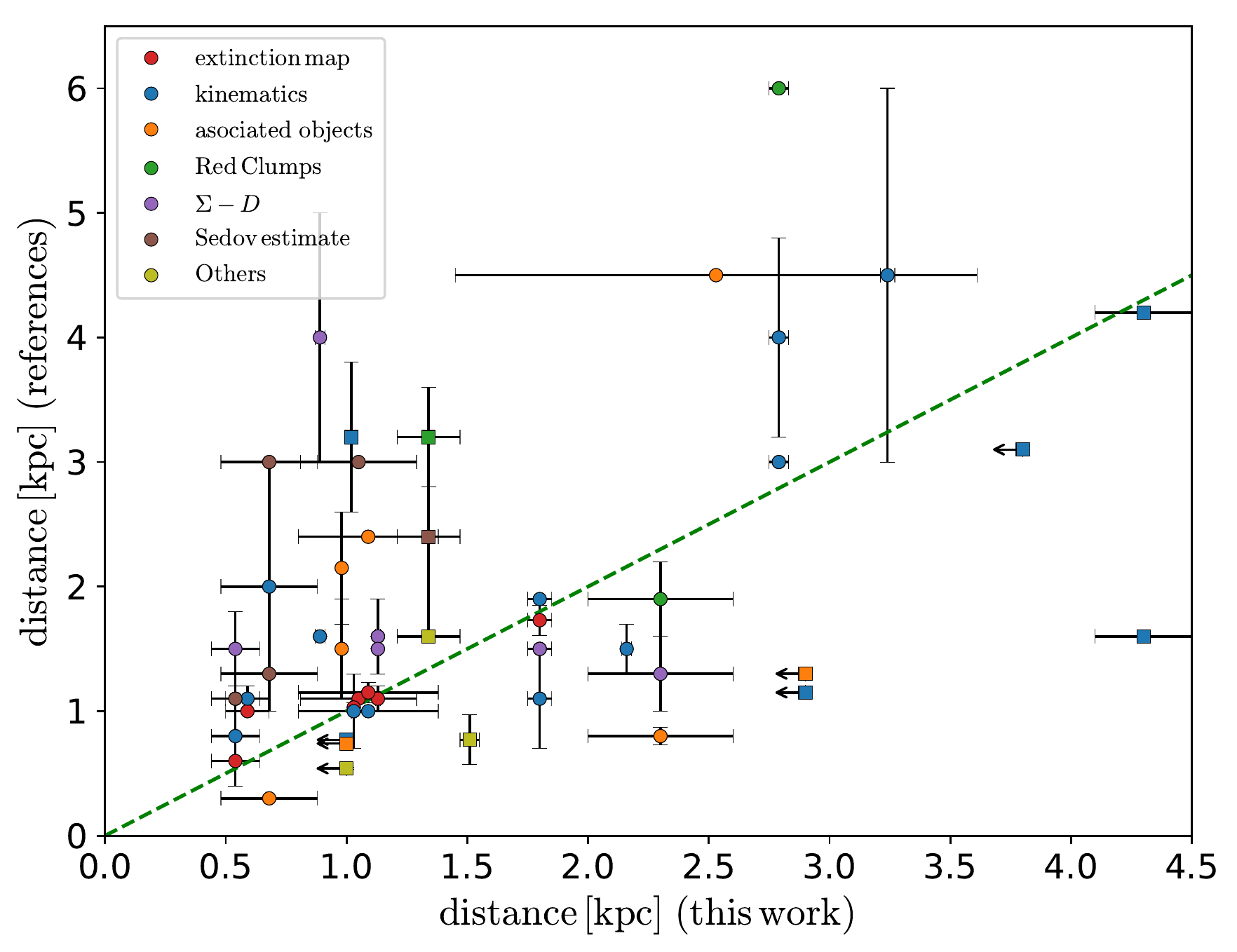}
   \caption{The comparison between our results and the distances measured by
   other methods for SNRs in Level A and B. Dots and squares decode SNRs
   in Level A and B respectively. Squares with left arrows are the cases with
   distance upper limits.
   Colors stand for different methods illustrated in the legends. ``Others'' means
   the methods based on proper motion or shock velocity. Dots without error-bars along
   the
   y-axis means no error analysis in the corresponding literature.
   The green dashed line traces the one-to-one correspondence.}
   \label{fig:dist-comp}
\end{figure}

\begin{figure}[ht]
   \centering
   \includegraphics[width=16.8cm]{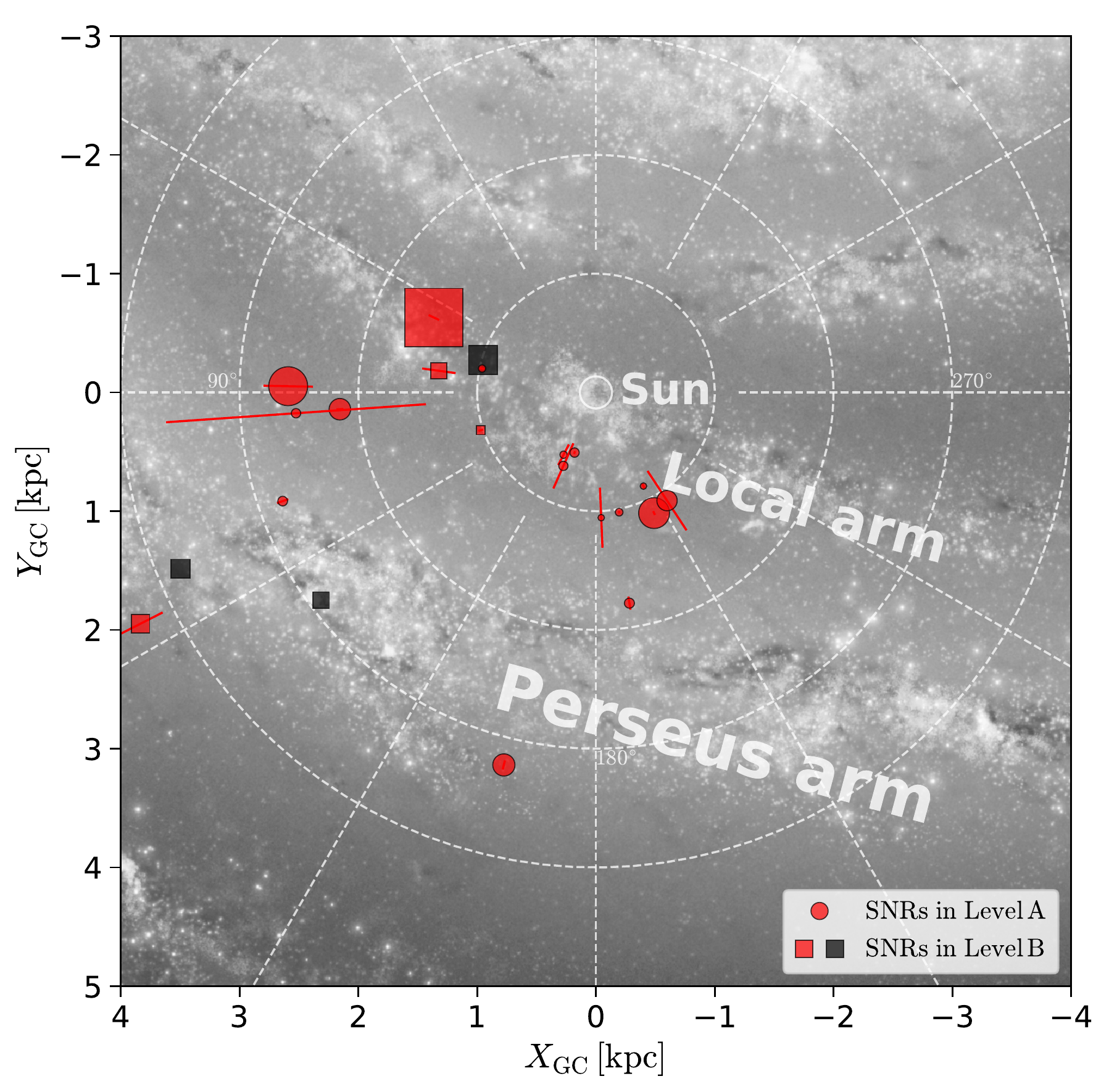}
   \caption{A face-on view of the distribution of the SNRs in Level A (15)
   and B (7). Red dots and squares with
   errorbars show the measured distances and uncertainties. Black squares are
   the cases with distance upper limits. The sizes are proportional to their
   diameters.
   The background image, created by Robert Hurt in consultation with
   Robert Benjamin \citep{Churchwell09}, is centered at the Sun, with the Galactic
   Center in the upward direction.}
   \label{fig:snr-distribution}
\end{figure}

\begin{figure}[ht]
  \centering
  \includegraphics[width=16.8cm]{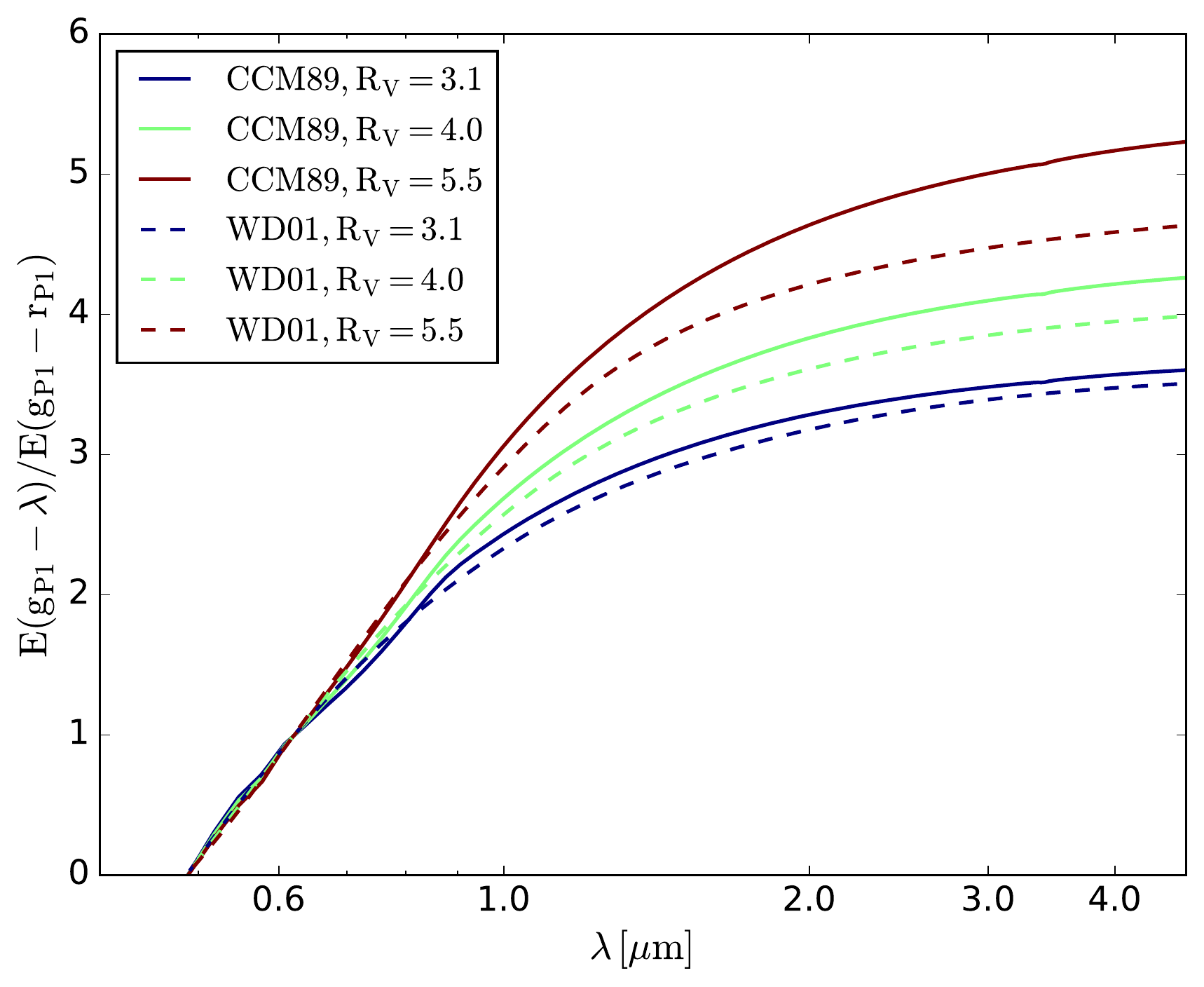}
  \caption{The comparison between the extinction curves
  calculated by CCM89 and WD01, with $\Rv=3.1$, 4.0, and 5.5.}
  \label{fig:wd01}
\end{figure}

\begin{figure*}[ht]
   \centering
   \includegraphics[width=16.8cm]{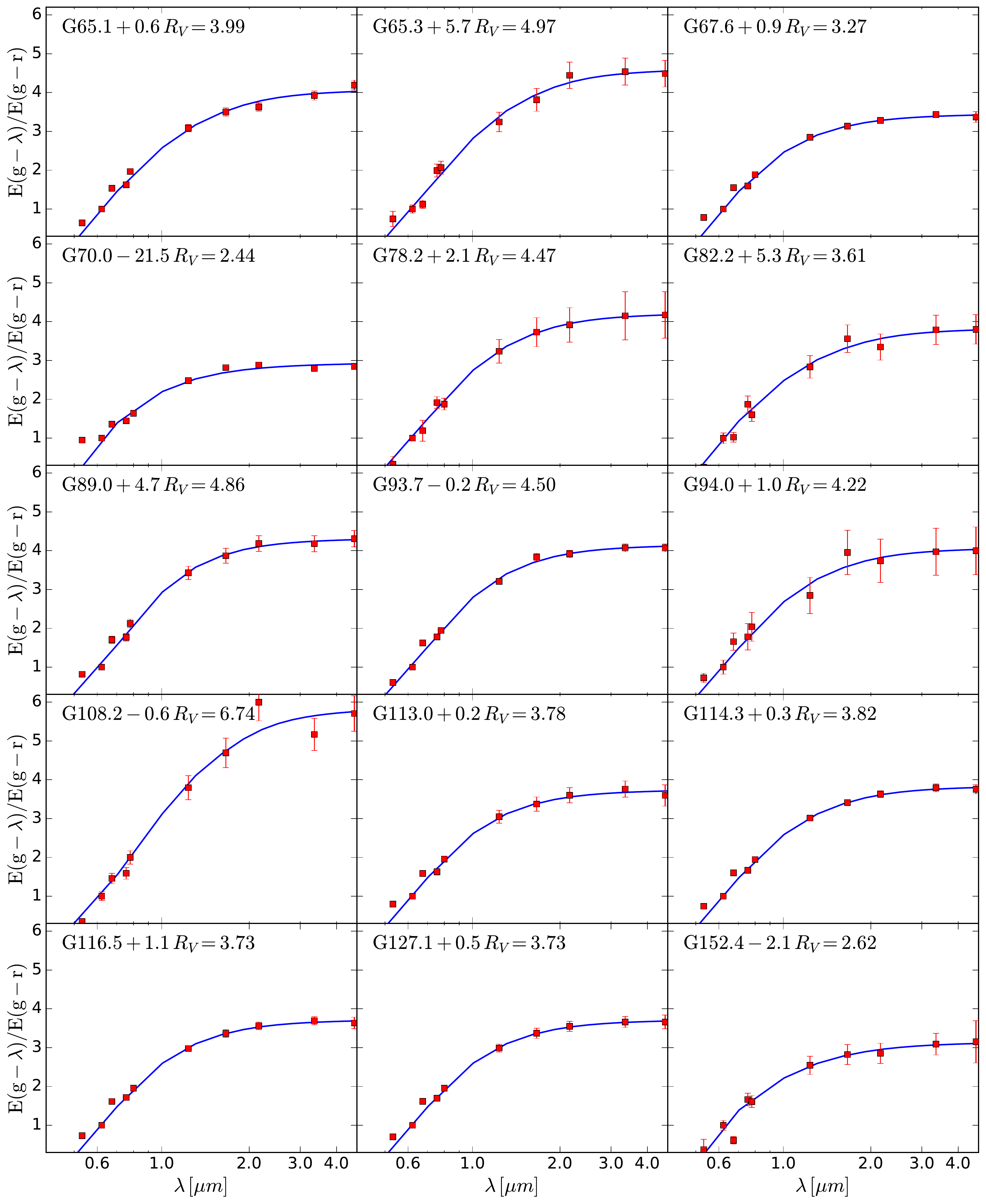}
   \caption{The extinction curves for 25 SNRs, as well as ``$l165$'' and Rosette
   Nebula. The red squares with errorbars are the CERs calculated for 10 bands,
   which from right to left are $\Gbp$, $\rps$, $G$, $\ips$, $\Grp$, $J$, $H$,
   $\Ks$, $W_1$, and $W_2$, respectively. The blue lines are the best-fit results
   of our dust model.}
   \label{fig:mrn-fit}
\end{figure*}\addtocounter{figure}{-1}

\begin{figure}[ht]
   \centering
   \includegraphics[width=16.8cm]{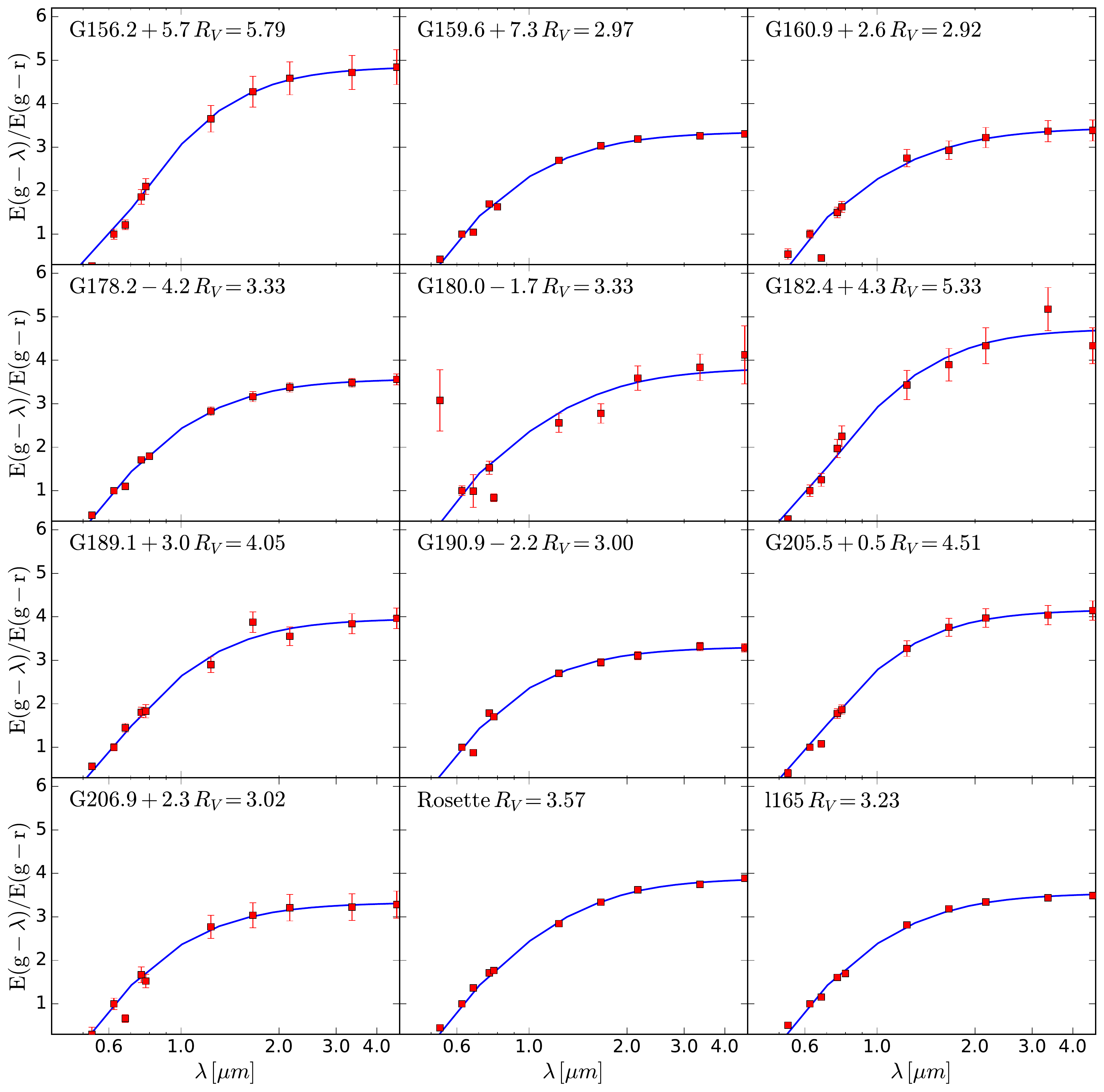}
   \caption{--continued.}
\end{figure}

\begin{figure}[!ht]
  \centering
  \includegraphics[width=16.8cm]{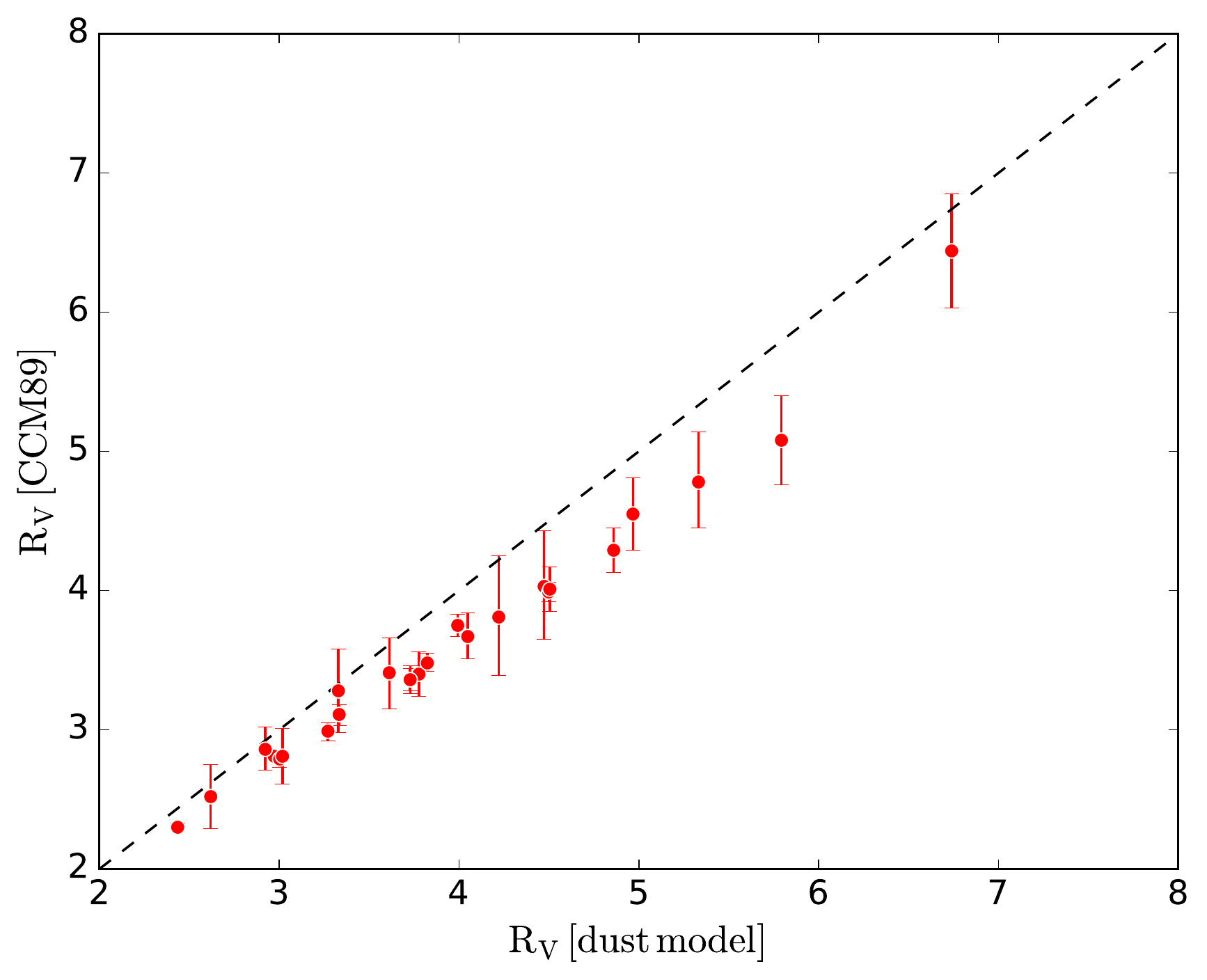}
  \caption{The comparison between $\Rv$ values fitted by the CCM89 formula
  and the dust model. The black dashed line traces the one-to-one correspondance.}
  \label{fig:mrn-ccm}
\end{figure}

\begin{figure}[ht]
  \centering
  \includegraphics[width=16.8cm]{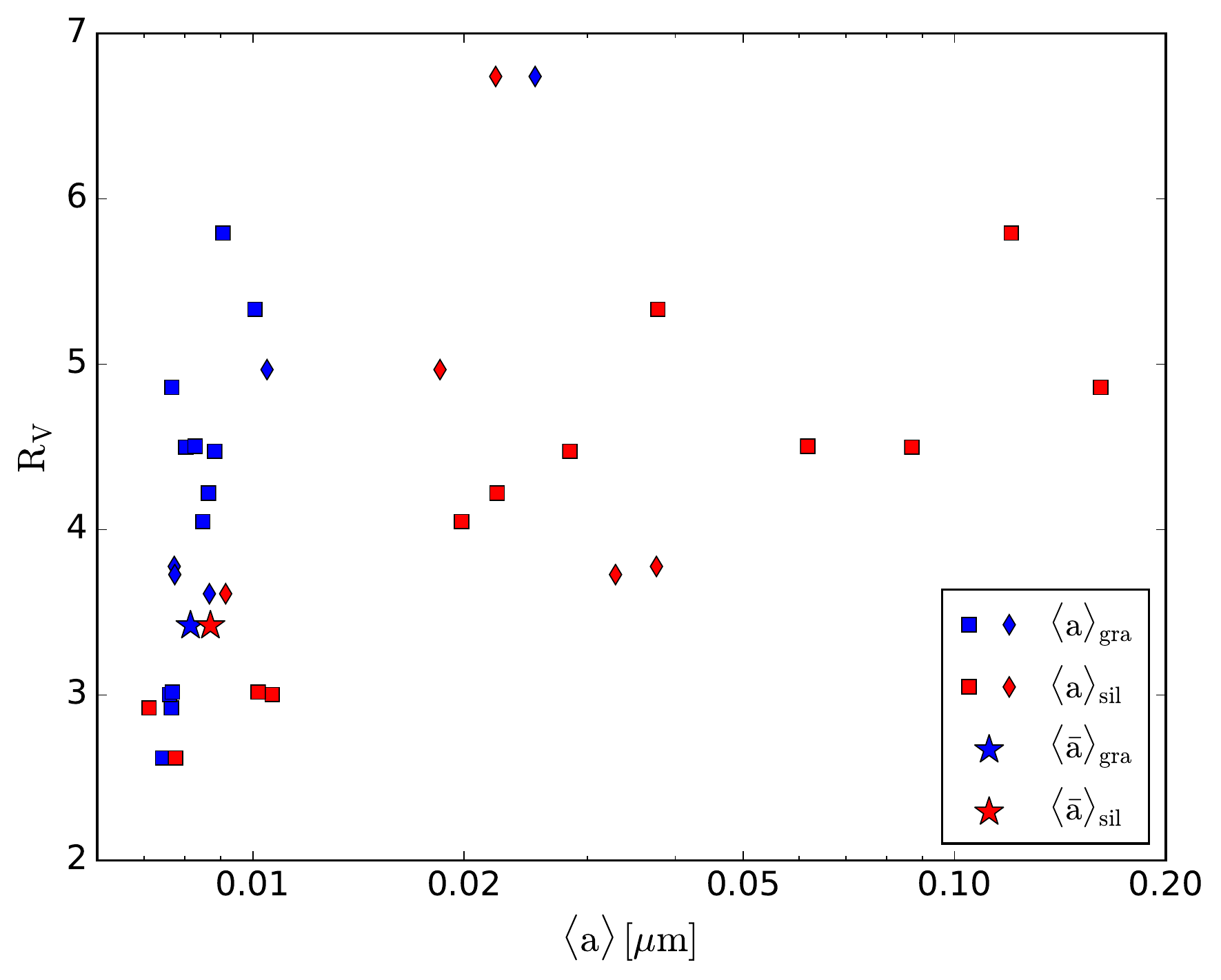}
  \caption{$\Rv$ values fitted by the dust model as a function of the
  average radii of graphite (in blue) and silicate (in red) grains respectively
  for 22 SNRs in Level A (squares) and B (diamonds). The big stars indicate
  the average of our Galaxy (see Section \ref{subsec:ave-curve}.)}
  \label{fig:mrn-com2}
\end{figure}

\begin{figure}[ht]
  \centering
  \includegraphics[width=16.8cm]{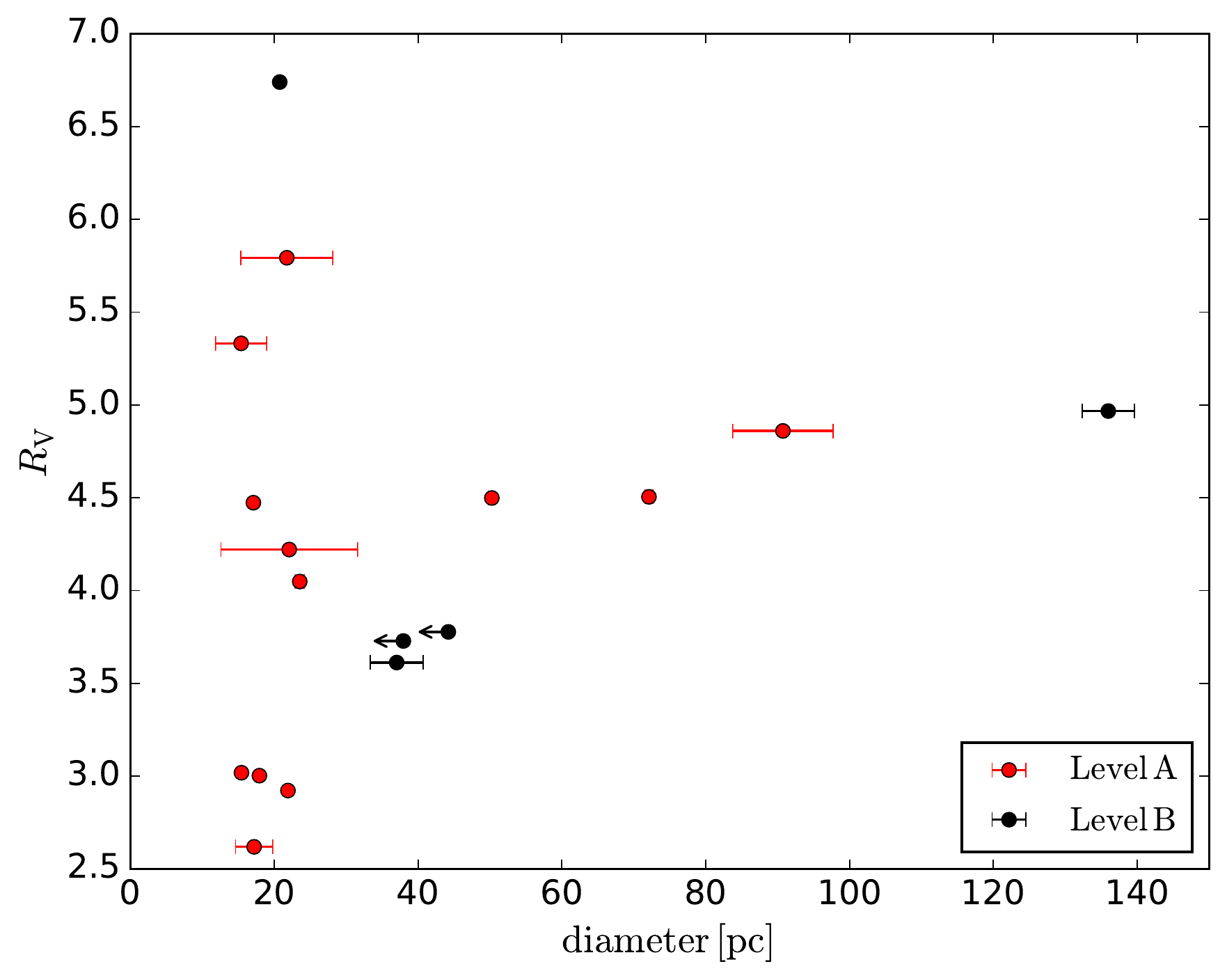}
  \caption{$\Rv$ values fitted by dust model as a function of the diameters
  of 22 SNRs in Level A (red dots) and B (black dots). The
  black dots with left arrows are the ones with upper distance limits.}
  \label{fig:mrn-com3}
\end{figure}

\begin{figure}[ht]
  \centering
  \includegraphics[width=16.8cm]{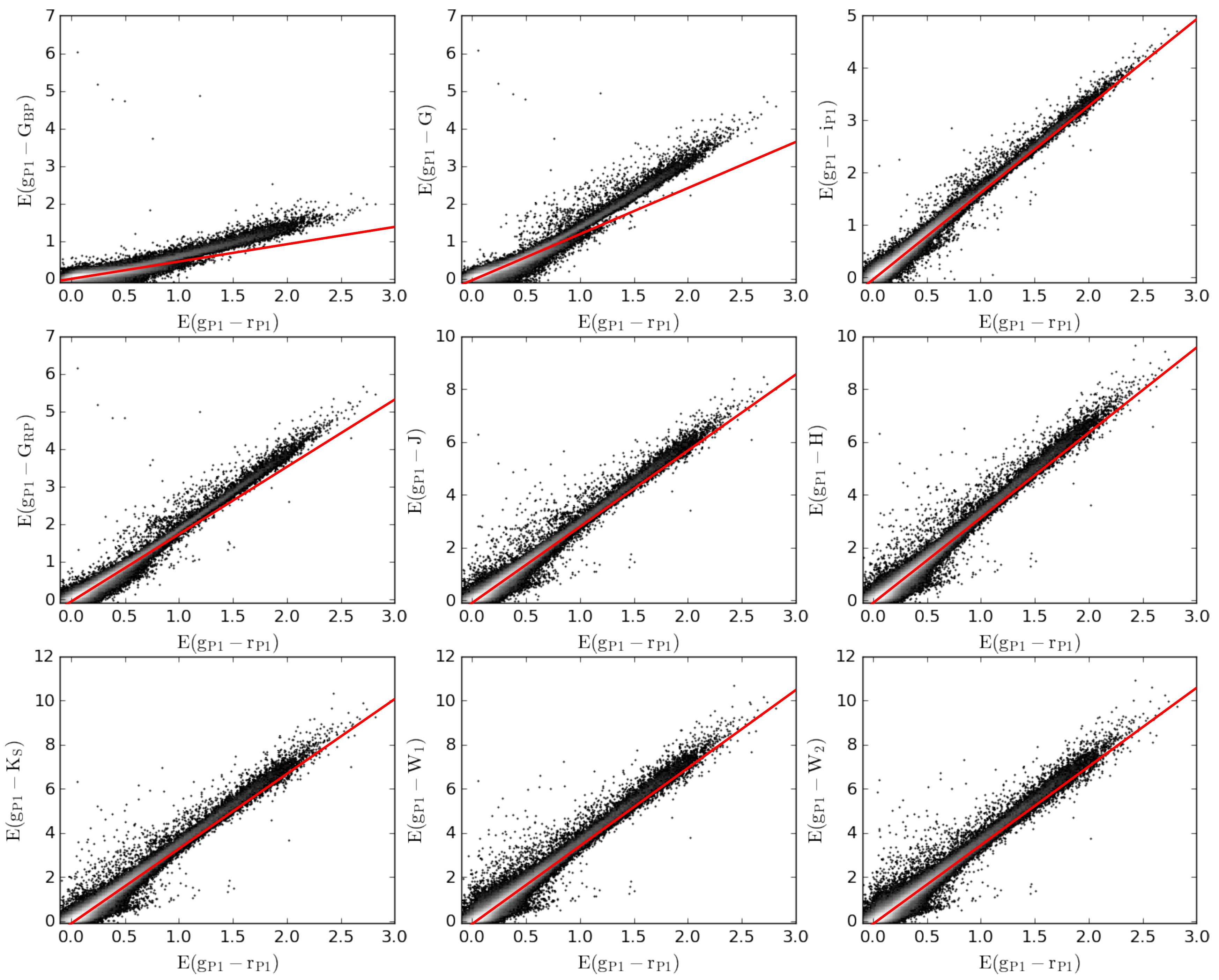}
  \caption{The color excess--color excess diagrams for our star sample. Red
  lines represent the best linear CERs fittings, and the grey-scale map shows the
  number density of the sample stars.}
  \label{fig:ave-cer}
\end{figure}

\begin{figure}[ht]
  \centering
  \includegraphics[width=16.8cm]{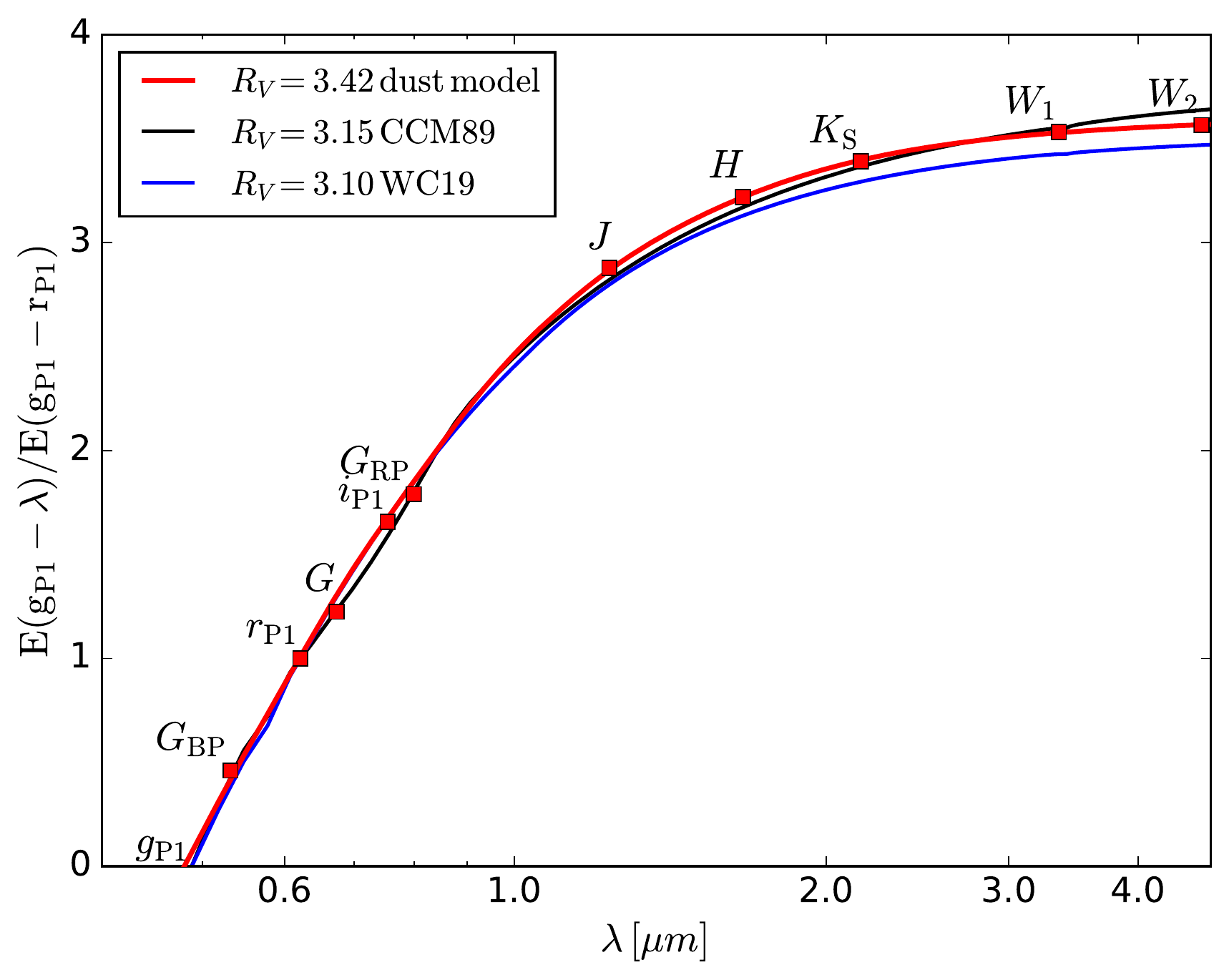}
  \caption{The average extinction curve of the Milky Way derived from our star sample.
  The
  red squares are calculated CERs at 10 bands, and the black line is fitted by CCM89.
  And the
  blue line is a modified curve suggested by \citet{Wang2019} with $\Rv=3.1$. The red
  line
  is fitted by our dust model. The {\it Gaia} bands, $\Gbp$, $G$, and $\Grp$, are
  presented,
  but not considered during fitting.}
  \label{fig:ext-curve}
\end{figure}

\end{document}